\documentclass[11pt]{book}

\usepackage[dvips]{epsfig}
\usepackage{amssymb}
\usepackage{amsbsy}
\usepackage{pstricks}
\usepackage{bm}
\usepackage{stmaryrd}

\catcode`\@=11

 \def\@normalsize{\@setsize\normalsize{13pt}\xipt\@xipt
   \abovedisplayskip 11pt plus3pt minus6pt
   \belowdisplayskip \abovedisplayskip
   \abovedisplayshortskip \z@ plus3pt
   \belowdisplayshortskip 6.6pt plus3.5pt minus3pt}

 \def\small{\@setsize\small{12pt}\xipt\@xipt
   \abovedisplayskip 10pt plus2pt minus5pt
   \belowdisplayskip \abovedisplayskip
   \abovedisplayshortskip \z@ plus3pt
   \belowdisplayshortskip 6pt plus3pt minus3pt
   \def\@listi{\topsep 6pt plus 2pt minus 2pt
     \parsep 3pt plus 2pt minus 1pt
     \itemsep \parsep}}

 \def\footnotesize{\@setsize\footnotesize{10pt}\ixpt\@ixpt
   \abovedisplayskip 8pt plus 2pt minus 4pt
   \belowdisplayskip \abovedisplayskip
   \abovedisplayshortskip \z@ plus 1pt
   \belowdisplayshortskip 4pt plus 2pt minus 2pt
   \def\@listi{\topsep 4pt plus 2pt minus 2pt
      \parsep 2pt plus 1pt minus 1pt
      \itemsep \parsep}}

 \def\scriptsize{\@setsize\scriptsize{9.5pt}\viiipt\@viiipt}
 \def\tiny{\@setsize\tiny{7pt}\vipt\@vipt}
 \def\large{\@setsize\large{14pt}\xiipt\@xiipt}
 \def\Large{\@setsize\Large{18pt}\xivpt\@xivpt}
 \def\LARGE{\@setsize\LARGE{22pt}\xviipt\@xviipt}
 \def\huge{\@setsize\huge{25pt}\xxpt\@xxpt}
 \def\Huge{\@setsize\Huge{30pt}\xxvpt\@xxvpt}

\def\section{\@startsection {section}{1}{\z@}%
{-1.5\baselineskip plus-1pt minus-3pt}{1\baselineskip plus1pt minus2pt}%
{\centering\normalsize\bf}}
\def\subsection{\@startsection{subsection}{2}{\z@}%
{-1\baselineskip plus-1pt minus-2pt}{1\baselineskip plus1pt minus2pt}%
{\normalsize\sc\noindent}}
\def\subsubsection{\@startsection{subsubsection}{3}{\z@}%
{-1\baselineskip plus-1pt minus-2pt}{1sp}{\normalsize\it\noindent}}
\def\paragraph{\@startsection{paragraph}{4}{\z@}%
{1\baselineskip plus1pt minus2pt}{-1em}{\normalsize\it\noindent}}
\let\subparagraph=\paragraph

\def\tableofcontents{\@restonecolfalse\if@twocolumn\@restonecoltrue
\onecolumn\fi\OSIDcont\@starttoc{con}\if@restonecol\twocolumn\fi}

\def\l@section{\@dottedtocline{1}{0em}{.66em}}

\def\thebibliography#1{\section*{{Bibliography}\@mkboth
 {BIBLIOGRAPHY}{BIBLIOGRAPHY}}\footnotesize\rm\list
 {[\arabic{enumi}]}{\settowidth\labelwidth{[#1]}\leftmargin\labelwidth
 \advance\leftmargin\labelsep\usecounter{enumi}}
 \def\newblock{\hskip .11em plus .33em minus -.07em}
 \sloppy\clubpenalty4000\widowpenalty4000
 \sfcode`\.=1000\relax}

\def\ps@myheadings{\let\@mkboth\@gobbletwo
\def\@oddhead{\hfil{\footnotesize\rm\rightmark}\hfil}
\def\@evenhead{\hfil{\footnotesize\rm\leftmark}\hfil}
\def\@oddfoot{\hfil{\footnotesize\sf\thepage}\hfil}
\def\@evenfoot{\hfil{\footnotesize\sf\thepage}\hfil}
\def\sectionmark##1{}\def\subsectionmark##1{}}

\def\@copyrighthead{}

\def\artid{0000001}
\def\Year{2008}        %
\newcounter{paPer}     %
\def\EndpagE{\expandafter\pageref{\the\value{paPer}OpSy}}

\def\ps@osiD{\let\@mkboth\@gobbletwo
\def\@oddhead{\@copyrighthead}
  \def\@oddfoot{\hfil{\footnotesize\sf\thepage}\hfil}
  \def\@evenhead{}\let\@evenfoot\@oddfoot}

\def\cite{\@ifnextchar [{\@tempswatrue\@Rcitex}{\@tempswafalse\@Rcitex[]}}

\def\@Rcitex[#1]#2{\if@filesw\immediate\write\@auxout{\string\citation{#2}}\fi
  \def\@citea{}\@cite{\@for\@citeb:=#2\do
    {\@citea\def\@citea{,\penalty\@m\,}\@ifundefined
       {b@\the\value{paPer}R\@citeb}{{\bf ?}\@warning
       {Citation `\@citeb' on page \thepage \space undefined}}%
\hbox{\csname b@\the\value{paPer}R\@citeb\endcsname}}}{#1}}

\long\def\@caption#1[#2]#3{\par\addcontentsline{\csname
  ext@#1\endcsname}{#1}{\protect\numberline{\csname
  the#1\endcsname}{\ignorespaces #2}}\begingroup
    \@parboxrestore
    \small                                        
    \@makecaption{\csname fnum@#1\endcsname}{\ignorespaces #3}\par
  \endgroup}

\newtoks\@stequation

\def\subequations{\refstepcounter{equation}%
\edef\@savedequation{\the\c@equation}%
\@stequation=\expandafter{\theequation}
\edef\@savedtheequation{\the\@stequation}
\edef\oldtheequation{\theequation}%
\setcounter{equation}{0}%
\def\theequation{\oldtheequation\alph{equation}}}%

\def\endsubequations{%
\setcounter{equation}{\@savedequation}%
\@stequation=\expandafter{\@savedtheequation}%
\edef\theequation{\the\@stequation}\global\@ignoretrue}

\catcode`\@=12

\let\Rlabel=\label
\let\Rbibitem=\bibitem
\let\Rref=\ref
\let\Rpageref=\pageref
\def\label#1{\expandafter\Rlabel{\the\value{paPer}R#1}}
\def\bibitem#1{\expandafter\Rbibitem{\the\value{paPer}R#1}}
\def\ref#1{\expandafter\Rref{\the\value{paPer}R#1}}
\def\pageref#1{\expandafter\Rpageref{\the\value{paPer}R#1}}

\def\thesection{\arabic{section}.}

\def\YYMm{\rule{0ex}{4em}}
\newtoks\TITsi
\newtoks\TITsii

\def\title#1{\def\TITs{\LARGE{\raggedright\noindent\YYMm #1%
\vskip8pt\par}}}

\def\author#1{\autMM{#1}\def\LHD{#1}}
\def\and{{\rm\lowercase{and}}}

\def\autMM#1{\TITsii={\vskip10pt\par\normalsize\rm\noindent #1\par}%
\TITsi=\expandafter{\TITs}\edef\TITs{\the\TITsi\the\TITsii}}

\def\address#1{\TITsii={\vskip6pt\par\footnotesize\sl\noindent #1\par}%
\TITsi=\expandafter{\TITs}%
\edef\TITs{\the\TITsi\the\TITsii}}

\def\received#1{\TITsii={\vskip10pt\par\small\rm\noindent(Received: #1)\par}%
\TITsi=\expandafter{\TITs}\edef\TITs{\the\TITsi\the\TITsii}}

\def\headtitle#1{\def\RHD{#1}}
\def\headauthor#1{\def\LHD{#1}}
\def\listas#1#2{\addcontentsline{con}{section}{{\sc #1: }{\rm #2}}}

\def\abst{{\bf Abstract.}}
\def\abstract#1{\TITs
       \vskip15pt\par\noindent
       {\footnotesize{\abst~} #1\vskip3pt\par}
       \markright{\RHD}
       \markboth{\LHD}{\RHD}}

\def\startpaper{%
       \cleardoublepage
       \setcounter{section}{0}
       \stepcounter{paPer}
       \setcounter{equation}{0}
       \setcounter{footnote}{0}
       \setcounter{figure}{0}
       \setcounter{table}{0}
       \def\theequation{\arabic{equation}}
       \def\thefootnote{\arabic{footnote}}
       \setcounter{defn}{0}
       \setcounter{thm}{0}
       \setcounter{lem}{0}
       \setcounter{prop}{0}
       \setcounter{rem}{0}
       \thispagestyle{osiD}}

\def\OSIDcont{\cleardoublepage\thispagestyle{empty}
       \markright{}\markboth{}{}
       \normalsize\rm
       \hspace*{\fill}{\large\rm
         Contents of the Volume \Volume, Number \Number}\hspace*{\fill}
       \par\vspace{1.5em}
       \par\noindent}

\def\endpaper{\expandafter\label{\the\value{paPer}OpSy}}

\def\1{{\mathchoice{\rm 1\mskip-4mu l}{\rm 1\mskip-4mu l}%
{\rm 1\mskip-4.5mu l}{\rm 1\mskip-5mu l}}}

\def\varkappa{\mbox{\bBB\char 123}}

\def\longhookrightarrow{\lhook\joinrel\relbar\joinrel\rightarrow}

\def\longhookUp{\lower6pt\hbox{\rotatebox{90}{$\longhookrightarrow$}}}

\def\tr{\mathop{\rm tr}}

\newtheorem{thm}{\rm THEOREM}

\newtheorem{lem}{\rm LEMMA}

\newtheorem{prop}{\rm PROPOSITION}

\newtheorem{defn}{\rm DEFINITION}
\newtheorem{exmp}{\rm EXAMPLE}

\newenvironment{pf}{\par\noindent{\it Proof.}\hspace{0.5em}}
{\hspace*{\fill}$\Box$\par\smallskip\noindent}

\def\theequation{\thesection\arabic{equation}}

\def\Myskip{\setlength{\baselineskip}{13pt}}

\def\text#1{\quad\mbox{\rm  #1 }\quad}

\def\DOInumber{}

\input xy
\xyoption{all}

\InputIfFileExists{psfig.sty}{\typeout{^^Jpsfig.sty inputed...ok}}{\typeout{^^JWarning: psfig.sty could not be found.^^J}}

\begin{document}

\def\artid{0000001}
\def\Volume{15}
\def\Number{1}
\def\Year{2008}
\setcounter{page}{1}

\def\DOInumber{}

\startpaper

\newcommand{\Mn}{M_n(\mathbb{C})}
\newcommand{\Mk}{M_k(\mathbb{C})}
\newcommand{\id}{\mbox{id}}
\newcommand{\ot}{{\,\otimes\,}}
\newcommand{{\Cd}}{{\mathbb{C}^d}}
\newcommand{\sbsigma}{{\mbox{\scriptsize \boldmath $\sigma$}}}
\newcommand{\sbalpha}{{\mbox{\scriptsize \boldmath $\alpha$}}}
\newcommand{\sbbeta}{{\mbox{\scriptsize \boldmath $\beta$}}}
\newcommand{\bsigma}{{\mbox{\boldmath $\sigma$}}}
\newcommand{\balpha}{{\mbox{\boldmath $\alpha$}}}
\newcommand{\bbeta}{{\mbox{\boldmath $\beta$}}}
\newcommand{\bmu}{{\mbox{\boldmath $\mu$}}}
\newcommand{\bnu}{{\mbox{\boldmath $\nu$}}}
\newcommand{\ba}{{\mbox{\boldmath $a$}}}
\newcommand{\bb}{{\mbox{\boldmath $b$}}}
\newcommand{\sba}{{\mbox{\scriptsize \boldmath $a$}}}
\newcommand{\MD}{\mathfrak{D}}
\newcommand{\sbb}{{\mbox{\scriptsize \boldmath $b$}}}
\newcommand{\sbmu}{{\mbox{\scriptsize \boldmath $\mu$}}}
\newcommand{\sbnu}{{\mbox{\scriptsize \boldmath $\nu$}}}
\def\oper{{\mathchoice{\rm 1\mskip-4mu l}{\rm 1\mskip-4mu l}%
{\rm 1\mskip-4.5mu l}{\rm 1\mskip-5mu l}}}
\def\<{\langle}
\def\>{\rangle}
\def\theequation{\thesection\arabic{equation}}

\newcommand{\RM}{\mathbb{R}}
\newcommand{\ZM}{\mathbb{Z}}
\newcommand{\QM}{\mathbb{Q}}
\newcommand{\NM}{\mathbb{N}}
\newcommand{\CM}{\mathbb{C}}
\newcommand{\LSR}{\mathcal{L}_{\mathrm{SR}}}
\newcommand{\LS}{\mathcal{L}_{\mathrm{S}}}
\newcommand{\LR}{\mathcal{L}_{\mathrm{R}}}
\newcommand{\Lz}{\mathcal{L}_{0}}
\newcommand{\Lt}{\mathcal{L}}

\newcommand{\rev}[1]{{\color{red}#1}}
\renewcommand{\d}{\mathrm{d}}
\newcommand{\e}{\mathrm{e}}
\renewcommand{\i}{\mathrm{i}}
\newcommand{\rint}{\int\displaylimits}
\renewcommand{\Re}{\mathop{\mathrm{Re}}\nolimits}
\renewcommand{\Im}{\mathop{\mathrm{Im}}\nolimits}
\renewcommand{\tr}{\mathop{\mathrm{tr}}\nolimits}
\newcommand{\Ran}{\mathop{\mathrm{Ran}}}

\title{On the derivation of the GKLS equation for weakly coupled systems}
\author{Paolo Facchi$^{1,2}$, Marilena Ligab\`o$^{3}$ and Kazuya Yuasa$^{4}$}
\address{$^1$ Dipartimento di Fisica and MECENAS, Universit\`a di Bari, I-70126 Bari, Italy} 
\address{$^2$ INFN, Sezione di Bari, I-70126 Bari, Italy}
\address{$^3$ Dipartimento di Matematica, Universit\`a di Bari, I-70125  Bari, Italy} 
\address{$^4$ Department of Physics, Waseda University, Tokyo 169-8555, Japan} 
\headauthor{Paolo Facchi, Marilena Ligab\`o and Kazuya Yuasa}
\headtitle{On the derivation of the GKLS equation}
\received{\today}
\listas{Paolo Facchi, Marilena Ligab\`o and Kazuya Yuasa}{On the derivation of the GKLS equation}

\abstract{
We consider the reduced dynamics of a small quantum system in interaction with a reservoir when the initial state is factorized. We present a rigorous derivation of a GKLS master equation in the weak-coupling limit for a generic bath, which is not assumed to have a bosonic or fermionic nature, and 
whose  reference state is not necessarily thermal. The crucial assumption is a reservoir state endowed with  a mixing property:
the $n$-point connected correlation function of the interaction must be asymptotically bounded by the product of two-point functions (clustering property).}

\Myskip



\section{Introduction}\label{intro}
\setcounter{equation}{0}

The reduced dynamics of a small quantum system in contact with a reservoir
is generally described in terms of a master equation, engendering an irreversible Markovian evolution. This description turns out to be extremely accurate and is commonly used in the description of a vast number of diverse physical situations. 
Excellent introductions to this subject can be found in Refs.~\cite{ref:QuantumNoise,Alicki07,Breuer07}.

However, the evolution of the total system is unitary and is described by a Schr\"odinger equation, whose reduction to the small system gives a completely positive dynamics, which in general is not Markovian and exhibits memory effects.
 Therefore, a fundamental question is the following: under which conditions does one obtain a master equation as a reduction of the Schr\"odinger equation?

According to a
widely accepted lore, the physical and mathematical assumptions that
are required in order to derive such an equation are three:
i) the reservoir is much larger than the system, ii) the coupling between them is very
weak, and iii)  the initial conditions are in a factorized form (initial statistical independence).

Under these assumptions, the system has a negligible influence on the
reservoir and the global properties of the latter remain unaffected
during the evolution. In turns, this enables one to assume that the
reservoir is in an equilibrium state, e.g.\ in a thermal state.

Mathematically, one considers concurrently a weak coupling limit and a long time limit (\emph{van Hove's limit}) of the  reduced dynamics of the small system. This limit turns out to be an irreversible Markovian dynamics: a completely positive semigroup preserving the trace of the density matrix of the small system. The generator of this semigroup is given in a Gorini-Kossakowski-Lindblad-Sudarshan (GKLS) form~\cite{GKS76,Lind76}. 

The weak coupling limit and the derivation of the resulting irreversible Markovian dynamics goes back to the work of Pauli, Weisskopf-Wigner and van Hove~\cite{WW,ref:VanHoveLimit}. For a review see Refs.~\cite{ref:KuboTextbook,Haake}. 
In the mathematical literature it was studied by Davies in two seminal papers~\cite{Davies,Davies2}, see also Ref.~\cite{ref:SpohnReview,Alicki07}.

The purpose of this article is to give a rigorous derivation of a GKLS master equation for a \emph{general reservoir}: in particular the equilibrium  state of the reservoir is not necessarily thermal and the bath is  not assumed to have a bosonic/fermionic nature. We will show that these two common assumptions can be disposed of. The crucial property 
the reservoir must satisfy is instead a \emph{clustering property} that, roughly speaking, implies that for large times the $n$-point connected correlation function of the interaction 
is bounded by the product of two-point functions where at least one of them is taken at two 
nonconsecutive times (gap condition). See Definition~\ref{defn:gapcond}. 

This behavior is in fact related to a \emph{mixing property} of the bath, an assumption that in Refs.~\cite{factorize1,factorize2} was already argued---on physical ground---to be crucial in the derivation  of a master equation.
This can be better understood by looking at the standard case of a bosonic/fermionic bath in a thermal equilibrium state. Indeed, in such a case the  $n$-point correlation function  can be written exactly in terms of product of two-point  functions by means of the Wick theorem. Moreover, the gap condition holds since the thermal state of a bosonic/fermionic bath  is in fact \emph{strongly mixing}, that is, for  any bath observables $A$, $B$ and $C$, one gets
\begin{equation}
\lim_{t \to + \infty} \langle A B(t)C\rangle= \langle A C \rangle \langle B \rangle,
\end{equation}
where $B(t)$ is the  evolution at time $t$ of the observable $B$ and $\langle{}\cdots{}\rangle$ is the expectation with respect to the thermal state. See Ref.~\cite{BR}.

In this sense we can say that the clustering property is related to  the strongly mixing property; in fact it is a stronger requirement.
Notice  that for a general reservoir no finite-rank interaction can satisfy the clustering property. 
From a physical point of view this means that, in order to obtain a Markovian dynamics, the interaction cannot be too localized: it has to connect the system with an infinite number of states of the reservoir.

A final remark is in order. In Refs.~\cite{factorize1,factorize2} the question of a correlated initial condition was also addressed, and it was argued that in that case too a mixing property of the reservoir is sufficient to get a GKLS equation in van Hove's limit. It would be interesting to understand whether the strategy of the proof used in this paper might also be applied to this more general situation.

The paper is organized as follows. In Sec.~\ref{sec:framework} we introduce some notation, set up the general framework of van Hove's limit,  and  introduce Nakajima-Zwanzig's projection operators and Davies' spectral average. 
In Sec.~\ref{sec:daviesreview} we review the abstract result on Banach spaces of Davies on the derivation of the master equation for the reduced dynamics of a system in van Hove's  limit (Lemma~\ref{lem:Dyson}  and Theorem~\ref{theorem:davies}). 
In Sec.~\ref{sec:diagram} we give an exact combinatorial formula for each term of the Dyson series of the reduced evolution in the coupling constant $\lambda$, and provide a diagrammatic expansion of each $n$-point correlation function (Theorem~\ref{thm:Kncomb}).  With this exact formula  we can introduce the clustering property, Definition~\ref{defn:gapcond}, as a sufficient assumption to control the convergence of the series. 
In Sec.~\ref{sec:mainth} we consider a class of quantum systems that satisfy the assumptions of the abstract Theorem~\ref{theorem:davies}, and thus yield a quantum dynamical semigroup in van Hove's limit (Theorem~\ref{thm:mainth}). 
In particular, in Proposition~\ref{lemmaKn} we prove that the Dyson series is norm convergent and in Proposition~\ref{lemmaKneps} we prove that each term of the series vanishes as $\lambda \to 0$. 
Finally, the Appendix contains a technical Lemma 
needed in the proof of Proposition~\ref{lemmaKneps}.

\section{Framework and notation}
 \setcounter{equation}{0}
 \label{sec:framework}
 
 We assume that the total system consists of a ``large'' reservoir $R$
and a ``small'' (sub)system $S$.

Let $(\mathfrak{M},S_R,\tau)$ be the quantum dynamical system of the reservoir, namely, $\mathfrak{M}$ is the algebra of the observables on $R$, $t \in \RM \mapsto \tau^t$ is a weakly continuous group of automorphism on $\mathfrak{M}$, and $S_R$ is an invariant faithful state; let $(\mathcal{H}_{R}, \pi ,\Omega_R)$ be the canonical cyclic representation of $\mathfrak{M}$ associated with $S_R$. The two conditions
\begin{equation}
\label{eq:GNS}
H_R \Omega_R=0 \text{and} \pi(\tau^t(A))=e^{itH_R}\pi(A)e^{-itH_R} \text{for all $A \in \mathfrak{M}$}
\end{equation}
uniquely determine a self-adjoint operator $H_R$ on the Hilbert space $\mathcal{H}_{R}$~\cite{ref:Aschbacher}.

Let $\mathcal{H}_{S}$ be the finite-dimensional Hilbert space of  the system $S$.
The total Hilbert space
$\mathcal{H}$ can be expressed as the tensor product of
the Hilbert spaces of the reservoir $\mathcal{H}_{R}$ and of
the system $\mathcal{H}_{S}$, namely
$\mathcal{H} =\mathcal{H}_S\otimes\mathcal{H}_R$.

The Hamiltonian of the total system is given by
\begin{equation}
H = H_0+ \lambda H_{SR} = H_{S}\otimes 1_{R} + 1_{S}\otimes H_{R}+\lambda W\otimes V , \label{Hamiltonian1}
\end{equation}
where $H_0=H_{S}\otimes 1_{R} + 1_{S}\otimes H_{R}$ is the free Hamiltonian of the total system, $H_S$ and $W$ are self-adjoint operators on $\mathcal{H}_S$,  $H_R$ and $V$ are self-adjoint operators on $\mathcal{H}_R$, and $\lambda \in \RM$ is the coupling constant. Moreover, we will always assume that $V$ is a bounded operator.

In order to describe the dynamics of the system at the level of density operators we introduce the Banach spaces  $\mathcal{T}(\mathcal{H})$, $\mathcal{T}(\mathcal{H}_S)$ and $\mathcal{T}(\mathcal{H}_R)$ of the trace class operators on $\mathcal{H}$, $\mathcal{H}_S$ and $\mathcal{H}_R$, respectively, and the Liouvillian of the total system
\begin{equation}
\mathcal{L}= \mathcal{L}_0 + \lambda \mathcal{L}_{SR},
\label{Liou1}
\end{equation}
where 
\begin{equation}
\mathcal{L}_0
=\mathcal{L}_{S}+\mathcal{L}_{R}
\end{equation}
is the free Liouvillian, describing the free uncoupled evolutions of
the system ($\mathcal{L}_{S}$) and of the reservoir
($\mathcal{L}_{R}$).
The domain of the Liouvillian $\mathcal{L}$ is given by all $\rho \in \mathcal{T}(\mathcal{H})$ such that  $\rho D(H) \subset D(H)$,
where $D(H)$ is the domain of the Hamiltonian $H$, and the action of the Liouvillian is $\mathcal{L}\rho =\mathcal{L}_0 \rho+ \lambda \mathcal{L}_{SR} \rho$, where
\begin{equation}
\mathcal{L}_0 \rho=[H_0, \rho] \text{and} \mathcal{L}_{SR} \rho=[W\otimes V, \rho].
\end{equation}
We define also the following operators
$$
\underline{\mathcal{L}}_S \sigma :=[H_S,\sigma] \text{and} \underline{\mathcal{L}}_R \omega :=[H_R,\omega]
$$
for all $\sigma \in \mathcal{T}(\mathcal{H}_S)$ and $\omega \in \mathcal{T}(\mathcal{H}_R)$ such that $\omega D(H_R) \subset D(H_R)$, where  $D(H_R)$ is the domain of the Hamiltonian $H_R$.

The evolution of  the total system is given by a group of isometries on $\mathcal{T}(\mathcal{H})$:
\begin{equation}
\rho_0 \mapsto \rho(t)=e^{-itH}\rho_0e^{itH} = e^{-it\mathcal{L}}\rho_0.
\end{equation}
The state of the system  $\sigma(t)$ at time $t$ is given by
\begin{equation}\label{eqn:sigtrace}
\sigma(t)=\tr_{R}(\rho(t))
\end{equation}
where $\tr_{R}:
\mathcal{T}(\mathcal{H})\to\mathcal{T}(\mathcal{H}_S)$
is the partial trace over the reservoir degrees of freedom. In general, unlike
$\rho(t)$, $\sigma(t)$ is not
unitarily equivalent to $\sigma(0)=\sigma_0$, and the system undergoes
dissipation and/or decoherence.
We are interested in the reduced dynamics of the system $S$, $\sigma(t)$ given by (\ref{eqn:sigtrace}).

Moreover, in general, due to memory effects, the reduced dynamics is \emph{not} given by a semigroup and does not satisfy a master equation. However, under suitable assumptions, one can obtain a quantum dynamical semigroup as a  limit of the above evolution.
The remarkable idea, proposed by van Hove in 1955~\cite{ref:VanHoveLimit}, is to consider a weaker and weaker interaction acting for a longer and longer time, that is the limit
\begin{equation}
\lambda \to 0, \text{keeping} \tau= \lambda^2 t  \text{finite.}
\label{eq:vHlim}
\end{equation}
One then looks at the reduced evolution (in the interaction picture) as a function of the rescaled (macroscopic) time $\tau$. This is called van Hove's ``$\lambda^2 t$'' limit and provides a rigorous justification of the Fermi ``golden'' rule~\cite{Fermi} and of the Weisskopf-Wigner approximation in quantum mechanics~\cite{WW}.

The procedure is the following. Let $P_{\Omega_R}$ be the rank-one projection associated with  the cyclic vector $\Omega_R \in \mathcal{H}_R$ in~(\ref{eq:GNS}). Then, $\omega_R =P_{\Omega_R}\in\mathcal{T}(\mathcal{H}_R)$ is the \textit{reference state} of the reservoir. Consider a factorized initial condition of the form
\begin{equation}
\label{eqn:initialcond}
\rho_0 = \sigma_0 \otimes \omega_R,
\end{equation}
where $\sigma_0 \in \mathcal{T}(\mathcal{H}_S)$ is an \emph{arbitrary} initial state of the system, i.e.\ $\sigma_0 \geq 0$, $\tr(\sigma_0)=1$. Notice that the stationarity in~(\ref{eq:GNS})
with respect to the reservoir
free dynamics reads
\begin{equation}
\underline{\mathcal{L}}_R \omega_R=0. 
\label{eqn:stationarity}
\end{equation}

Our aim is to prove that, under suitable assumptions, van Hove's limit
\begin{equation}
\label{eqn:limit dynamics}
\sigma_{I}(\tau) =\lim_{\lambda\to 0} e^{i\frac{\tau}{\lambda^2} \underline{\mathcal{L}}_S}\tr_R\!\left(e^{-i \frac{\tau}{\lambda^2} (\mathcal{L}_0+\lambda \mathcal{L}_{SR}) }(\sigma_0 \otimes \omega_R) \right)
\end{equation}
exists for all $\sigma_0\in \mathcal{T}(\mathcal{H}_S)$ and for all $\tau\geq 0$, and that $\sigma_{I}(\tau)$ is the solution of a master equation
\begin{equation}
\label{eqn:reduced dynamics}
\frac{\d}{\d \tau}\sigma_{I}(\tau) = -\mathcal{K}\sigma_{I}(\tau) , \qquad
\sigma_I(0)= \sigma_0 ,
\end{equation}
where $\mathcal{K}$ is a GKLS generator  acting on the (finite-dimensional) Banach space $\mathcal{T}(\mathcal{H}_S)$.

A useful tool  will be  Nakajima-Zwanzig's projection operators~\cite{ref:Projection,ref:Zwanzig,ref:KuboTextbook}
\begin{equation}
\label{eqn:gendefproj}
P\rho=\tr_{R}(\rho)\otimes \omega_R
=\sigma\otimes\omega_R,\qquad
Q=1-P,
\end{equation}
where $\rho\in\mathcal{T}(\mathcal{H})$ and
$\sigma\in\mathcal{T}(\mathcal{H}_S)$.
Note that, from the normalization condition $\tr_{R}(\omega_R)=1$,
it follows that $P^2=P$,
$Q^2=Q$ and $PQ=QP=0$. Therefore, $P$ is the
projection onto the space $P\mathcal{T}(\mathcal{H})$, whose elements have the form
$\sigma\otimes\omega_R$. Thus, $P\mathcal{T}(\mathcal{H})$ is a finite-dimensional subspace of
$\mathcal{T}(\mathcal{H})$ isometrically isomorphic
to~$\mathcal{T}(\mathcal{H}_S)$.

We immediately get that
\begin{equation}
[P,\mathcal{L}_S]
=[Q,\mathcal{L}_{S}]=0,\qquad
e^{-i t \mathcal{L}_{R}} P = P e^{-i t \mathcal{L}_{R}} = P.
\label{eqn:propPQ}
\end{equation}
The first equation is a consequence of the fact that
$\mathcal{L}_{S}$ and $P$ essentially operate in
different spaces, while the second derives from (\ref{eqn:stationarity}) and
from the characteristic structure of the Liouvillians,
$\tr(\mathcal{L}\rho)=0$ (a direct consequence of probability
conservation). In addition, we require that
\begin{equation}\label{eqn:propPQ3}
P\mathcal{L}_{SR}P=0,
\end{equation}
which, for  a nonconstant $W$, is equivalent to the condition 
\begin{equation}
\tr (V\omega_R)=0.
\label{eq:meanV0}
\end{equation}

By making use of (\ref{eqn:propPQ}) and (\ref{eqn:propPQ3}), the
total Liouvillian can be formally decomposed as
\begin{equation}
\mathcal{L}
=P\mathcal{L}_{S}P
+Q\mathcal{L}_RQ +\lambda
Q\mathcal{L}_{SR}Q +\lambda
P\mathcal{L}_{SR}Q +\lambda
Q\mathcal{L}_{SR}P.
\label{eqn:Liouvdec}
\end{equation}
Therefore, the free evolutions generated by $\mathcal{L}_{S}$ and
$\mathcal{L}_{R}$ leave invariant the two subspaces
$\Ran P$ and $\Ran Q$, and all
transitions are driven by the interaction~$\mathcal{L}_{SR}$.

Finally, let us introduce a  device that will be useful later.
Let us consider the spectral decomposition of the Hamiltonian $H_S$ of the system~$S$:
\begin{equation}
H_S=\sum_{j}\varepsilon_j P_j.
\end{equation}
It induces a spectral decomposition of the corresponding Liouvillian~${\mathcal{L}}_S$,
\begin{equation}
{\mathcal{L}}_S = \sum_{\alpha} \omega_{\alpha}  Q_{\alpha},
\end{equation}
where 
\begin{equation}
Q_\alpha \rho:=\sum_{j,k}\delta_{\omega_\alpha,\varepsilon_j -\varepsilon_k } (P_j\otimes 1_R) \rho (P_k \otimes 1_R) ,
\end{equation}
for all $\rho \in \mathcal{T}(\mathcal{H})$, and $\omega_{\alpha}$ are distinct and real, representing all possible energy gaps of the free system $S$. 
It is immediate to check that $Q_\alpha Q_\beta= \delta_{\alpha,\beta}Q_\alpha$, so $\{Q_\alpha\}_{\alpha}$ is a family of projections, and one gets
\begin{equation}
e^{-it {\mathcal{L}}_S}=\sum_{\alpha}e^{-it\omega_{\alpha}} Q_{\alpha}.
\end{equation}

Given a bounded operator $X:\mathcal{T}(\mathcal{H} ) \to \mathcal{T}(\mathcal{H} )$, we define its \emph{spectral average} as~\cite{Davies}
\begin{equation}\label{eqn:natural}
X^{\natural}=\sum_{\alpha}Q_{\alpha}XQ_{\alpha},
\end{equation}
which can be easily proved to be equivalent to
\begin{equation}\label{eqn:natural2}
X^{\natural}=\lim_{T \to +\infty} \frac{1}{T} \int_{0}^{T}  e^{it {\mathcal{L}}_S}Xe^{-it {\mathcal{L}}_S} \, \d t,
\end{equation}
an expression that makes no reference to the spectral projections $\{Q_{\alpha}\}_{\alpha}$. 

We will see that the spectral average will turn a bounded operator on density matrices into a GKLS generator, a crucial ingredient for having a completely positive Markovian evolution.

\section{A review of Davies' results}\label{sec:daviesreview}
In this section we recall the abstract result of Davies~\cite{Davies} on the derivation of the master equation for the reduced dynamics of the system (in the interaction picture) in  van Hove's  limit~(\ref{eq:vHlim}), namely, when both the weak-coupling limit ($\lambda \to 0$) and the long-time limit ($t= \tau / \lambda^2 \to +\infty$) are considered.

Let $\rho_0 = \sigma_0 \otimes \omega_R \in P\mathcal{T}(\mathcal{H})$ be the initial state as in~(\ref{eqn:initialcond}). Consider the system $S$ in the interaction picture at  van Hove's time scale $t= \tau / \lambda^2$. In order to prove the existence of the limit reduced dynamics $\sigma_I(\tau)$ in (\ref{eqn:limit dynamics}), we will study instead the following limit on the full space $\mathcal{T}(\mathcal{H})$,
\begin{equation}
\rho_I(\tau) = \sigma_I(\tau) \otimes \omega_R=\lim_{\lambda \to 0} U^{\lambda}(\tau)\rho_0,
\end{equation}
where
\begin{equation}
U^{\lambda}(\tau)=e^{i\frac{\tau}{\lambda^2} \mathcal{L}_S} Pe^{-i\frac{\tau}{\lambda^2} (\mathcal{L}_0+\lambda \mathcal{L}_{SR})}P.
\end{equation}
As discussed above, this problem is equivalent to~(\ref{eqn:limit dynamics}), since the spaces $\mathcal{T}(\mathcal{H}_S)$ and $P\mathcal{T}(\mathcal{H})$ are isometrically isomorphic.

First of all, we establish, in an abstract setting, an integral equation for $U^{\lambda}(\tau)$ and give a series representation for its kernel. This will be the starting point of all the following investigation. 

Notice that all the results of this section are valid in an \emph{abstract} Banach space $\mathcal{B}$. However, with an abuse of notation, we will keep denoting the abstract operators by the physical notation discussed above, so that the reader, by looking at theorems, can immediately understand where we are aiming at.

\begin{lem}
\label{lem:Dyson}
Let $P=P^2$ be a finite-rank projection on a Banach space $\mathcal{B}$. Let $t \mapsto e^{-i t \mathcal{L}_{R}}$ be a strongly continuous group of isometries on $\mathcal{B}$, which commutes with $P$ and acts as the identity on $\Ran P$:
\begin{equation}
e^{-i t \mathcal{L}_{R}} P = P e^{-i t \mathcal{L}_{R}} = P.
\end{equation}
Let $\mathcal{L}_{S}$ and $\mathcal{L}_{SR}$   be bounded operators such that 
$-i\mathcal{L}_0 = -i(\mathcal{L}_{R} + \mathcal{L}_{S})$ and $-i(\mathcal{L}_0 +\lambda \mathcal{L}_{SR})$ are the generators of strongly continuous groups of isometries on $\mathcal{B}$, and
\begin{equation}
\mathcal{L}_{S} = P \mathcal{L}_S P, \qquad P\mathcal{L}_{SR}P=0.
\end{equation}
For any $\lambda, \tau \in \RM$, let
\begin{equation}
U^{\lambda}(\tau)=e^{i\frac{\tau}{\lambda^2} \mathcal{L}_S} Pe^{-i\frac{\tau}{\lambda^2} (\mathcal{L}_0+\lambda \mathcal{L}_{SR})}P.
\label{eq:Ulambdadef}
\end{equation}
Then, $U^{\lambda}(\tau)$ satisfies
\begin{equation}\label{eqn:Ylambdataus}
U^{\lambda}(\tau)=P-\int_{0}^{\tau} e^{i \frac{u}{\lambda^2} \mathcal{L}_S}K^\lambda(\tau-u)e^{-i \frac{u}{\lambda^2} \mathcal{L}_S}U^{\lambda}(u) \, \d u,
\end{equation}
where
\begin{equation}\label{eqn:Klambdatau}
K^\lambda(\tau)=\int_{0}^{\tau/\lambda^2}Pe^{is \mathcal{L}_0}\mathcal{L}_{SR}Qe^{-is(\mathcal{L}_0+\lambda Q\mathcal{L}_{SR}Q)}Q\mathcal{L}_{SR}P \, \d s.
\end{equation}
Moreover, $K^\lambda(\tau)$ can be given by the norm  convergent series
\begin{equation}
K^\lambda(\tau) =  \int_{0}^{\tau/\lambda^2} P\mathcal{L}_{SR}(s)Q\mathcal{L}_{SR}P\, \d s +  \sum_{n=1}^{+\infty}(-i\lambda)^n K_n(\tau/\lambda^2),
\label{eq:Klambdaseries}
\end{equation}
where $\mathcal{L}_{SR}(s) =e^{is\mathcal{L}_0}\mathcal{L}_{SR}e^{-is \mathcal{L}_0}$, 
\begin{equation}\label{defn:Kntabstract}
K_{n}(t)=\int_{\Delta^{n+1}(t)} P \mathcal{L}_{SR}(z_{n+1})Q \mathcal{L}_{SR}(z_n)Q\cdots Q\mathcal{L}_{SR}(z_1)Q\mathcal{L}_{SR}P\, \d z,
\end{equation}
and
\begin{equation}
\Delta^{n+1}(t)=\{z=(z_1,\dots,z_{n+1})\in\RM^{n+1} \,|\, 0\leq z_1 \leq \cdots \leq z_{n+1}\leq t\}
\end{equation}
is the $(n+1)$-dimensional simplex.
\end{lem}
\begin{pf}
Set $A= -i (\mathcal{L}_0 + \lambda Q \mathcal{L}_{SR} Q)$ and $B= -i \lambda (P \mathcal{L}_{SR} Q + Q \mathcal{L}_{SR} P)$, so that $A +B = -i (\mathcal{L}_0 + \lambda \mathcal{L}_{SR})$. Thus,  
$$
U^{\lambda}(\tau) = e^{-t A} P e^{t(A+B)}P,
$$
with $t= \tau/\lambda^2$, because $e^{-t A} P =  e^{i t \mathcal{L}_S} P$.
Since $B$ is a bounded perturbation, the group  of isometries $t\mapsto e^{t A}$ and $t\mapsto e^{t(A+B)}$ are related by Dyson's equation
\begin{equation}
e^{t(A+B)} = e^{t A} + \int_0^t e^{(t-s) A} B e^{s(A+B)}\,\d s ,
\label{eq:Dysonint}
\end{equation}
where the integral is in the strong topology~\cite{Engel00}. By iterating,
\begin{eqnarray*}
\hspace*{-5truemm}&&
e^{t(A+B)}=e^{tA} 
+ \int_0^t e^{(t-s) A} B e^{s A}\, \d s 
\\
\hspace*{-5truemm}&&\hphantom{e^{t(A+B)}=e^{tA}}
{} + \int_0^t \left(\int_0^{t-u}  e^{(t-u-s)A} B e^{s A} B \,\d s \right)e^{u(A+B)}\, \d u  .
\end{eqnarray*}
Since $e^{t A} P = P e^{t A}$ and $PBP=0$, one has
$$
e^{-t A}  P e^{t(A+B)} P = P + \int_0^t e^{-u A} \left(\int_0^{t-u} P e^{-s A} B e^{s A} B P\, \d s \right) P e^{u(A+B)} P\, \d u .
$$
Therefore, by plugging the definitions of $A$, $B$ and $t$, and by a change of integration variable, we have~(\ref{eqn:Ylambdataus}) and~(\ref{eqn:Klambdatau}).

Set now $A= -i \mathcal{L}_0$ and $B= -i \lambda Q \mathcal{L}_{SR} Q$, so that $A +B = -i (\mathcal{L}_0 + \lambda Q\mathcal{L}_{SR}Q)$. Equation~(\ref{eq:Dysonint}) holds, and by iterating it we get Dyson's series,
\begin{eqnarray*}
\hspace*{-5truemm}&&
e^{-is(\mathcal{L}_0+\lambda Q\mathcal{L}_{SR}Q)}
\\
\hspace*{-5truemm}&&
\qquad
=e^{-is\mathcal{L}_0}
+e^{-is\mathcal{L}_0} \sum_{n=1}^{+\infty}(-i\lambda)^n \int_{\Delta^{n}(s)} Q\mathcal{L}_{SR}(z_{n})Q \cdots Q\mathcal{L}_{SR}(z_1)Q\,\d z,
\end{eqnarray*}
which plugged into~(\ref{eqn:Klambdatau}) gives~(\ref{eq:Klambdaseries}).
\end{pf}

The following theorem contains the result in Ref.~\cite{Davies} concerning the limit of $U^\lambda(\tau)$ for  $\lambda \to 0$. We consider a small variation of the original theorem, which is convenient for our later discussion.

\begin{thm}\label{theorem:davies}
Under the assumptions of Lemma~\ref{lem:Dyson}, suppose that the  operator 
\begin{equation}\label{eqn:def_K}
\bm{K}:=\int_{0}^{+\infty}  P\mathcal{L}_{SR}(s)Q \mathcal{L}_{SR}P \,\d s 
\end{equation}
on the Banach space $\mathcal{B}$ is well defined, namely that
\begin{equation}\label{eqn:existence K}
\int_{0}^{+\infty} \| P\mathcal{L}_{SR}(s)Q \mathcal{L}_{SR}P\| \,\d s  < +\infty.
\end{equation}
Suppose that there exists a sequence $(c_n)_{n \geq 1}$ such that the power series
$$
\sum_{n\geq 1}c_n s^n
$$
has infinite radius of convergence and  
\begin{equation}
\label{eqn:convK}
\| K_{n}(t) \| \leq c_n t^{[n/2]}, \text{for all $n \geq 1$ and $t\geq 0$,}
\end{equation}
where $[x]$ denotes the integer part of $x$, i.e.\ the largest integer $\leq x$.
Suppose that for all $m \geq 1$   there exist  $d_{m} \geq 0$ such that for all $t > 0$
\begin{equation}
\label{eqn:convK_eps}
\| K_{2m}(t) \| \leq {d}_m t^{m-\epsilon},
\end{equation}
for some  $\epsilon >0$.

Then, one has
\begin{equation}\label{eqn:limiting dynamics}
\lim_{\lambda \to 0} U^{\lambda}(\tau)=e^{-\tau \bm{K}^{\natural}} P
\end{equation}
uniformly in $\tau \in [0, \tau_1]$, for all $\tau_1 >0 $, where 
\begin{equation}\label{eqn:natural4}
\bm{K}^{\natural}=\lim_{T \to +\infty} \frac{1}{T} \int_{0}^{T} e^{it\mathcal{L}_S} \bm{K} e^{-it\mathcal{L}_S} \, \d t 
\end{equation}
is Davies' spectral average of $\bm{K}$, which always exists since $\mathcal{L}_S$ has a finite (pure point) spectrum.
\end{thm}
\begin{pf}
\paragraph*{\emph{Step 1}.} Fix $\tau_0$ and $\tau_1$, with $0<\tau_0< \tau_1$. We prove that
\begin{equation}
\label{eqn:klambda to k}
\lim_{\lambda \to 0} K^\lambda(\tau)=\bm{K}
\end{equation}
uniformly in $\tau \in [\tau_0, \tau_1]$. We observe that
\begin{eqnarray*}
\hspace*{-7truemm}&&
\|K^\lambda(\tau)-\bm{K}\|
\leq\int_{\tau / \lambda^2}^{+\infty} \|P\mathcal{L}_{SR}(s)Q \mathcal{L}_{SR}P \| \, \d s
+ \sum_{n=1}^{+\infty}|\lambda|^{n}\|K_{n}(\tau/\lambda^2)\| 
\\
\hspace*{-7truemm}&&\hphantom{\|K^\lambda(\tau)-\bm{K}\|}
=\int_{\tau / \lambda^2}^{+\infty}\| P\mathcal{L}_{SR}(s)Q \mathcal{L}_{SR}P \| \, \d s + \sum_{m=1}^{+\infty}|\lambda|^{2m}\| K_{2m}(\tau/\lambda^2)\|\\
\hspace*{-7truemm}&&\hphantom{\|K^\lambda(\tau)-\bm{K}\|={}}
{}+\sum_{m=0}^{+\infty}|\lambda|^{2m+1}\| K_{2m+1}(\tau/\lambda^2) \| ,
\end{eqnarray*}
and using (\ref{eqn:existence K}) one gets
$$
\int_{\tau / \lambda^2}^{+\infty} \| P\mathcal{L}_{SR}(s)Q \mathcal{L}_{SR}P\| \, \d s \to 0
$$
as $\lambda \to 0$, uniformly in $\tau \in [\tau_0, \tau_1]$. Moreover, by using 
(\ref{eqn:convK}) and 
(\ref{eqn:convK_eps}), it is easy to check that 
$$
 \sum_{m=1}^{+\infty}|\lambda|^{2m}\| K_{2m}(\tau/\lambda^2) \|
$$
is a uniformly convergent series in $\tau \in [0, \tau_1]$, which vanishes term by term as $\lambda \to 0$. 
Finally, by using (\ref{eqn:convK}) we have that 
$$
\sum_{m=0}^{+\infty}|\lambda|^{2m+1}\| K_{2m+1}(\tau/\lambda^2) \| \leq |\lambda|  \sum_{m=0}^{+\infty} c_{2m+1} \tau^m  \to 0
$$ 
uniformly in $\tau \in [\tau_0, \tau_1]$ as $\lambda \to 0$.

\paragraph*{\emph{Step 2}.}
Let $\mathcal{V}:=C([0,\tau_1]; \Ran P)$. We claim that for all $\sigma \in \mathcal{V}$,
$$
\lim_{\lambda \to 0} \int_{0}^{\tau}    e^{i\frac{u}{\lambda^2} \mathcal{L}_S} K^\lambda(\tau-u) e^{-i\frac{u}{\lambda^2} \mathcal{L}_S} \sigma(u)  \, \d u=\int_{0}^{\tau}  \bm{K}^{\natural}\sigma(u)\, \d u
$$
uniformly in $\tau \in [0, \tau_1]$.
Indeed, by using (\ref{eqn:klambda to k}) it can be easily shown that
$$
\left\| \int_{0}^{\tau} e^{i\frac{u}{\lambda^2} \mathcal{L}_S} K^\lambda(\tau-u) e^{-i\frac{u}{\lambda^2} \mathcal{L}_S}  \sigma(u) \, \d u- \int_{0}^{\tau}  e^{i\frac{u}{\lambda^2} \mathcal{L}_S} \bm{K} e^{-i\frac{u}{\lambda^2} \mathcal{L}_S}  \sigma(u) \, \d u \right\| \to 0
$$
as $\lambda\to 0$, uniformly in $\tau \in [0, \tau_1]$. 

Moreover, since $\Ran P$ is finite-dimensional and $\mathcal{L}_S = P \mathcal{L}_S P$, one gets $e^{-i t \mathcal{L}_S} =\sum_{\alpha}  e^{-i t \omega_{\alpha}} Q_{\alpha}$, 
where $\{Q_\alpha\}$ are the spectral projections of $\mathcal{L} _S$ and $\{\omega_\alpha\}$ are the distinct eigenvalues. Therefore, we get
$\bm{K}^{\natural} = \sum_\alpha Q_\alpha \bm{K} Q_\alpha$, as in~(\ref{eqn:natural2}), whence
\begin{eqnarray*}
\hspace*{-5truemm}&&
\left\| \int_{0}^{\tau}  e^{i\frac{u}{\lambda^2} \mathcal{L}_S} \bm{K} e^{-i\frac{u}{\lambda^2} \mathcal{L}_S} \sigma(u)\, \d u - \int_{0}^{\tau}  \bm{K}^{\natural}\sigma(u) \, \d u  \right\| \nonumber\\
\hspace*{-5truemm}&&\qquad
 =   \left\| \int_{0}^{\tau}  \sum_{\alpha, \beta} Q_{\alpha}\bm{K}Q_{\beta} e^{i \frac{u}{\lambda^2} (\omega_\alpha - \omega_\beta)} \sigma(u)\, \d u  - \int_{0}^{\tau} \bm{K}^{\natural}  \sigma(u)\, \d u  \right\| \nonumber\\
\hspace*{-5truemm}&&\qquad
 \to   \left\| \int_{0}^{\tau} \sum_{\alpha,\beta} Q_{\alpha}\bm{K}Q_{\beta} \delta_{\omega_\alpha, \omega_\beta} \sigma(u) \, \d u - \int_{0}^{\tau}   \sum_{\alpha} Q_{\alpha}\bm{K}Q_{\alpha} \sigma(u)\, \d u  \right\| = 0
\end{eqnarray*}
as $\lambda\to0$, uniformly in $\tau \in [0, \tau_1]$.

\paragraph*{\emph{Step 3}.} Let $\rho_0 \in \Ran P$.
Define for all $\tau \in [0, \tau_1]$
\begin{equation}
\label{eqn:def sigmal and sigma}
\rho^{\lambda}(\tau)=U^{\lambda}(\tau) \rho_0 \text{and} \rho(\tau)=e^{-\tau \bm{K}^{\natural}} \rho_0.
\end{equation}
Of course, $\rho^{\lambda}({}\cdot{}), \rho({}\cdot{}) \in \mathcal{V}$.
We will prove that
$$
\lim_{\lambda \to 0}\rho^{\lambda}(\tau)=\rho(\tau)
$$
uniformly in $\tau \in [0, \tau_1]$.
It follows immediately by~(\ref{eqn:def sigmal and sigma}) and by Lemma~\ref{lem:Dyson} that
\begin{equation}
\label{Volterra}
\rho^{\lambda}(\tau)-\rho(\tau) =\sum_{n=1}^{+\infty}(-1)^n \int_{\Delta^{n}(\tau)} [A_n^{(\tau,\lambda)}(u) -(\bm{K}^{\natural})^n]\rho_0 \, \d u,
\end{equation}
where 
$$
A_n^{(\tau,\lambda)}(u) :=H^\lambda(\tau-u_n,u_n)H^\lambda(u_n-u_{n-1},u_{n-1})\cdots H^\lambda(u_{2}-u_1,u_1) ,
$$
with
$$
H^\lambda(\tau,u)=e^{i \frac{u}{\lambda^2}\mathcal{L}_S }K^\lambda(\tau)e^{-i \frac{u}{\lambda^2} \mathcal{L}_S }.
$$
Moreover  the series in (\ref{Volterra}) is dominated by a totally convergent series. Indeed,
$$
\left\| \int_{\Delta^{n}(\tau)}A_n^{(\tau,\lambda)}(u) \rho_0 \, \d u \right\| \leq \frac{1}{n!}(\|\bm{K}\|+c)^n\tau_1^n\|\rho_0\|
$$
with some $c\ge0$ for any $\lambda\le\lambda_0$ for a small enough $\lambda_0$, and
$$
\left\| \int_{\Delta^{n}(\tau)}(\bm{K}^{\natural})^n \rho_0 \, \d u \right\| \leq \frac{1}{n!}\|\bm{K}\|^n\tau_1^n\|\rho_0\|.
$$
Therefore,
$$
\sup_{\tau \in [0,\tau_1]}\|\rho^{\lambda}(\tau)-\rho(\tau)\| \leq  \sum_{n=1}^{+\infty}\frac{2}{n!}(\|\bm{K}\|+c)^n\tau_1^n\|\rho_0\|.
$$
Thus we have proved that each term of the series (\ref{Volterra}) vanishes as $\lambda \to 0$ uniformly in $\tau \in [0, \tau_1]$. Therefore the series converges to zero as $\lambda \to 0$, and this completes the proof. 
\end{pf}

\section{Diagrammatic expansions}\label{sec:diagram}

Now we go back to our problem and look in more detail at the structure of the operator $K^\lambda(t)$ given by~(\ref{eqn:Klambdatau})
in the  case of the Banach space $\mathcal{B}=\mathcal{T}(\mathcal{H})$ and with the operators introduced in Sec.~\ref{sec:framework}.
Our aim is to show that, under suitable conditions, our concrete realization satisfies the hypotheses of the abstract Theorem~\ref{theorem:davies}, and thus it gives rise to a quantum dynamical semigroup in van Hove's limit.

Let us gather here the assumptions on our model discussed in Sec.~\ref{sec:framework}.

\paragraph*{\emph{ASSUMPTIONS A:}}
\begin{enumerate}
\item Let  $\mathcal{H}_R$ be a complex separable Hilbert space, and $t\mapsto e^{-it H_R}$ be a unitary group, with self-adjoint generator $H_R$.
\item There exists a unit vector $\Omega_R \in \mathcal{H}_R$ which is invariant, namely $H_R \Omega_R =0$. Let $\omega_R = P_{\Omega_R}$ be the rank-one projection onto the span of $\Omega_R$.
\item Let $\mathcal{H}_S$ be a finite-dimensional complex Hilbert space, and $H_S$ a self-adjoint operator in $\mathcal{H}_S$. 
\item Let $W\otimes V$ be a bounded operator on the tensor product  $\mathcal{H} = \mathcal{H}_S\otimes \mathcal{H}_R$, with $W$ and $V$ self-adjoint, and with $\tr(V\omega_R)=0$.
\item Let  $P\rho=\tr_{R}(\rho)\otimes \omega_R$ and $Q=1-P$, for $\rho\in \mathcal{T}(\mathcal{H})$ be projection operators on the Banach space $\mathcal{B}=\mathcal{T}(\mathcal{H})$.
\item Let $t\mapsto e^{-i t \mathcal{L}_0}$ be the group of isometries on $\mathcal{B}$ defined by
$e^{-i t \mathcal{L}_0} \rho = e^{-i t (H_S\otimes 1_R + 1_S \otimes H_R)} \rho e^{i t (H_S\otimes 1_R + 1_S \otimes H_R)}$, and let $\mathcal{L}_{SR}\rho = [W\otimes V, \rho]$, for all $\rho\in\mathcal{B}$.
\end{enumerate}

Under these assumptions, the hypotheses of Lemma~\ref{lem:Dyson} are satisfied and the kernel of the evolution operator $U^\lambda (\tau)$ is given by the sum of the series~(\ref{eq:Klambdaseries}).  In this section we  aim at proving an exact formula and a diagrammatic expansion of the $n$-th term of the series, $K_n(t)$ given in~(\ref{defn:Kntabstract}).
This diagrammatic expansion will be crucial to prove our main theorem. In order to present the result we introduce some notation.

\subsection{Definitions, notations and examples}

\begin{defn}
Let $n \in \NM$, $n\geq 1$. 
\begin{enumerate}
\item We set $\llbracket n\rrbracket :=\{0,1,\dots,n\}$.
\item Let $A \subset \llbracket n+1\rrbracket $. We put $\bar{A}:= \llbracket n+1\rrbracket  \setminus A$ and we denote by $|A|$ the number of elements of $A$. 
\end{enumerate}
\end{defn}

\begin{defn}
Let $n \in \NM$, $n\geq 1$.  We define the set of noncrossing partitions of $\llbracket n\rrbracket$, and we denote it by $\mathrm{NC}_{n}$ the family of partitions of
 the sequence $(0,1,\dots,n)$ into contiguous subsequences of length larger than 1. In detail: 
 $d \in \mathrm{NC}_{n}$ if there exist $r \geq 1$ and $k_1, \dots, k_r \in \NM\setminus\{0,1\}$, $k_1 + \cdots +k_r=n+1$, such that $d=(d_1, \dots, d_r)$ where
$$
d_1=(0, \dots, k_1-1), d_2=(k_1, \dots, k_1+k_2-1), \dots, d_r=(k_1+\cdots+k_{r-1}, \dots,  n).
$$
We denote by $|d|=r$ the number of subsequences in $d$, and by $|d_j|=k_j$ the length of the subsequence $d_j$, for $j=1, \dots, r$.
\end{defn}
\begin{exmp}
Consider $\llbracket 7\rrbracket =\{0,1,2,3,4,5,6,7\}$. Two partitions in $\mathrm{NC}_7$ are
$$
d=((0,1),(2,3,4),(5,6,7)), \quad d'= ((0,1,2),(3,4,5),(6,7)).
$$
\end{exmp}

\begin{defn}\label{defn:dA}
Let $m \geq 1$ and $a=(a_1, \dots, a_m) \in \NM^m$, $a_j < a_{j+1}$. Let $A \subset \NM$. We consider the two disjoint sets 
$$
\{a_1, \dots, a_m\} \cap A = \{r_1, \dots,r_k\}
$$
 and
$$
\{ a_1, \dots, a_m\} \setminus  A = \{s_{k+1}, \dots,s_m\},
$$
and we assume that $r_1 <\cdots < r_k$ and $s_{k+1} >\cdots >s_m$. We define the rearrangement of $a$ by $A$ as the
$m$-tuple 
$$
a^A=(r_1,\dots,r_k,s_{k+1},\dots,s_m).
$$ 
\end{defn}
\begin{exmp}
Let $d=((0,1),(2,3,4),(5,6,7,8)) \in \mathrm{NC}_8$, with $d_1=(0,1)$, $d_2=(2,3,4)$, $d_3=(5,6,7,8)$, and let $A=\{ 1,3,5,6\}$. Then,
$$
d_1^A=(1,0), \quad
d_2^A=(3,4,2),   \quad
d_3^A=(5,6,8,7).
$$
\end{exmp}

\begin{defn}
Let $(F_k)_{k \in \NM}$ be a sequence of bounded operators in a Banach space.  We  define three different ordered products:
\begin{enumerate}
\item If $a=(a_1, \dots, a_m) \in \NM^m$,  with  $a_j \neq a_k$ for $j \neq k$, we denote the ordered product by
$$
\prod_{k \in a} F_k:= F_{a_1}F_{a_2} \cdots F_{a_m}. 
$$
\item 
If $A=\{j_1, j_2, \ldots, j_r\} \subset \NM$, with $j_1<j_2< \cdots <j_r$, then we set
$$
\mathop{\overrightarrow{\prod}}_{k \in A} F_k:= F_{j_1}F_{j_2} \cdots F_{j_r}  \text{and} \mathop{\overleftarrow{\prod}}_{k \in A} F_k:= F_{j_r}F_{j_{r-1}} \cdots F_{j_1}.
$$
\end{enumerate}
\end{defn}

\subsection{Diagrammatic expansion of $K_n(t)$}
Using the above notations and definitions we can present the following result.
\begin{thm}\label{thm:Kncomb}
If Assumptions A hold,   it results that the operator $K_n(t)$ defined in~(\ref{defn:Kntabstract}) has the following structure:
\begin{equation}\label{eqn:Kncomb}
K_n(t) \rho = \sum_{A \subset \llbracket n+1\rrbracket }  (-1)^{|A|} \int_{\Delta^{n+1}(t)}  \mathcal{G}_n(A,z) \mathop{\overleftarrow{\prod}}_{j \in \bar{A}}W(z_{j}) \sigma \mathop{\overrightarrow{\prod}}_{k \in A} W(z_k) \otimes \omega_R \, \d z , 
\end{equation}
for all $\rho \in \mathcal{T}(\mathcal{H})$, where  $\sigma=\tr_R(\rho)$ (namely $P\rho=\sigma \otimes \omega_R$), $z_0:=0$, and
\begin{equation}\label{eqn:GnA}
\mathcal{G}_n(A,z)= \sum_{d \in \mathrm{NC}_{n+1}} (-1)^{|d|+1} \prod_{s=1}^{|d|} \tr\!\left( \prod_{k \in d_{s}^A } V(z_k)\omega_R \right).
\end{equation}
See Figs.~\ref{fig1} and \ref{fig3} for the Feynman diagrams of (\ref{eqn:Kncomb}) and (\ref{eqn:GnA}).
\end{thm}
\begin{pf}
 Let us recall the definition of $K_n(t)$,
$$
K_{n}(t)=\int_{\Delta^{n+1}(t)} P \mathcal{L}_{SR}(z_{n+1})Q \mathcal{L}_{SR}(z_n)Q\cdots Q\mathcal{L}_{SR}(z_1)Q\mathcal{L}_{SR}(z_0)P\, \d z,
$$
where $z_0:=0$, and observe that
\begin{eqnarray}\label{eqn:prodPLQLQLP}
\hspace*{-5truemm}&&
P \mathcal{L}_{SR}(z_{n+1})Q \mathcal{L}_{SR}(z_n)Q\cdots Q\mathcal{L}_{SR}(z_1)Q\mathcal{L}_{SR}(z_0)P \nonumber \\
\hspace*{-5truemm}&&\qquad
 = P \mathcal{L}_{SR}(z_{n+1})(1-P) \mathcal{L}_{SR}(z_n)(1-P)\cdots (1-P)\mathcal{L}_{SR}(z_0)P.\nonumber\\ 
\hspace*{-5truemm}&&
\end{eqnarray}
The presence/absence of a projection $P$ in~(\ref{eqn:prodPLQLQLP}) splits the operator into a sum of many terms, each one of them being related to a specific partition of $n+2$, the total number of variables. 
Using this idea, it is not difficult to prove that $K_n(t)$ can be rewritten as follows:
\begin{eqnarray}
\hspace*{-10truemm}&&
P \mathcal{L}_{SR}(z_{n+1})Q \mathcal{L}_{SR}(z_n)Q\cdots Q\mathcal{L}_{SR}(z_1)Q\mathcal{L}_{SR}(z_0)P \nonumber \\
\hspace*{-10truemm}&&\qquad
 = \sum_{d \in \mathrm{NC}_{n+1}}(-1)^{|d|+1} \prod_{j\in (|d|, |d|-1, \dots, 1)}\left(P \prod_{k \in \tilde{d_j}}\mathcal{L}_{SR}(z_{k})P\right),
\label{comb1}
\end{eqnarray}
where $\tilde{d}_j$ is the reversed sequence of $d_j$, that is, if $d_j=(a_1, \dots, a_r)$, then $\tilde{d}_j=(a_r, \dots, a_1)$.
Observe that given $d=(d_1, \dots, d_k) \in \mathrm{NC}_{n+1}$, the length of $d_j$ represents the distance between two successive projections $P$, and this is the reason for the request in $\mathrm{NC}_{n+1}$  that $|d_j| \geq 2$ (because   $P\mathcal{L}_{SR}(z)P=0$ for all $z \in \RM$). Let us consider some examples of possible $d \in \mathrm{NC}_{n+1}$.
\begin{figure}
\centering
\resizebox{0.99\textwidth}{!}{\includegraphics{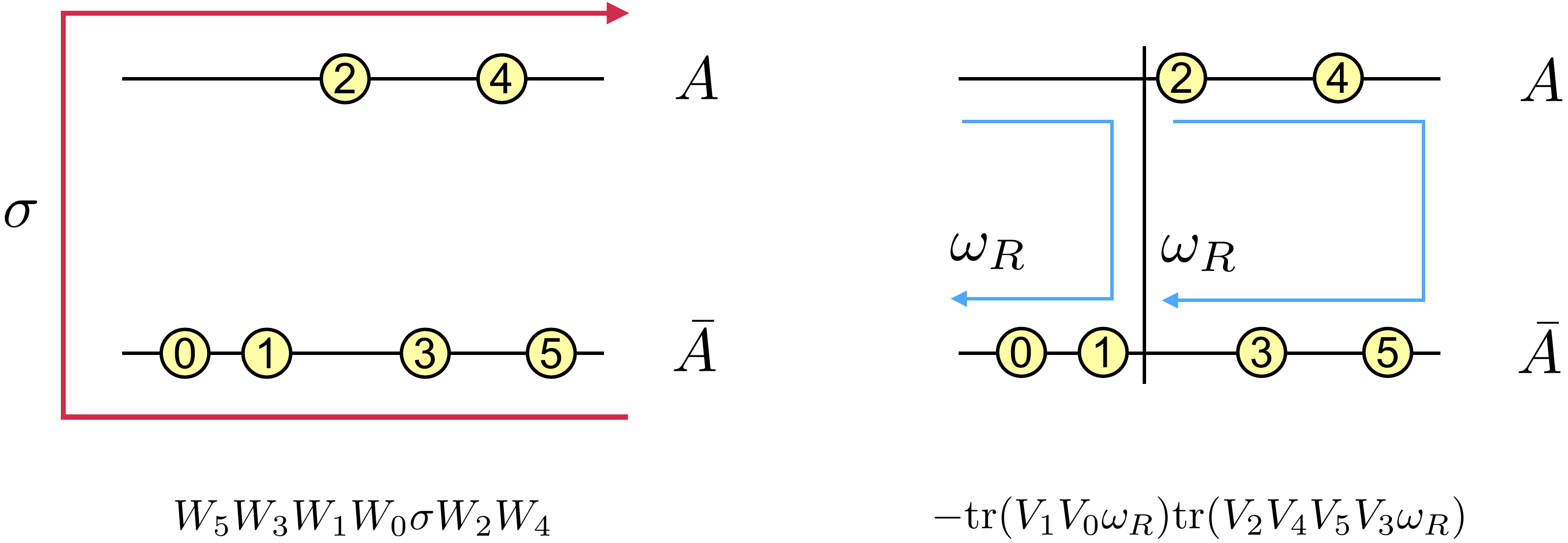}
}
\caption{(Color online) Left: Feynman diagram of the term $\prod_{j \in \bar{A}}^{\leftarrow}W(z_{j}) \sigma \prod_{k \in A}^{\rightarrow} W(z_k)$ for $n=4$ and $A=\{2,4\}$, where $W_l:=W(z_l)$. Right: Feynman diagram of the term $(-1)^{|d|+1}\prod_{s=1}^{|d|} \tr\!\left( \prod_{k \in d_{s}^A } V(z_k)\omega_R \right)$ for $n=4$, $A=\{2,4\}$, $d=(d_1,d_2)$, $|d|=2$, $d_1=(0,1)$, $d_2=(2,3,4,5)$, $d_1^A=(0,1)$, $d_2^A=(2,4,5,3)$, where $V_l:=V(z_l)$.}
\label{fig1} 
\end{figure}
\begin{figure}
\centering
\resizebox{0.99\textwidth}{!}{\includegraphics{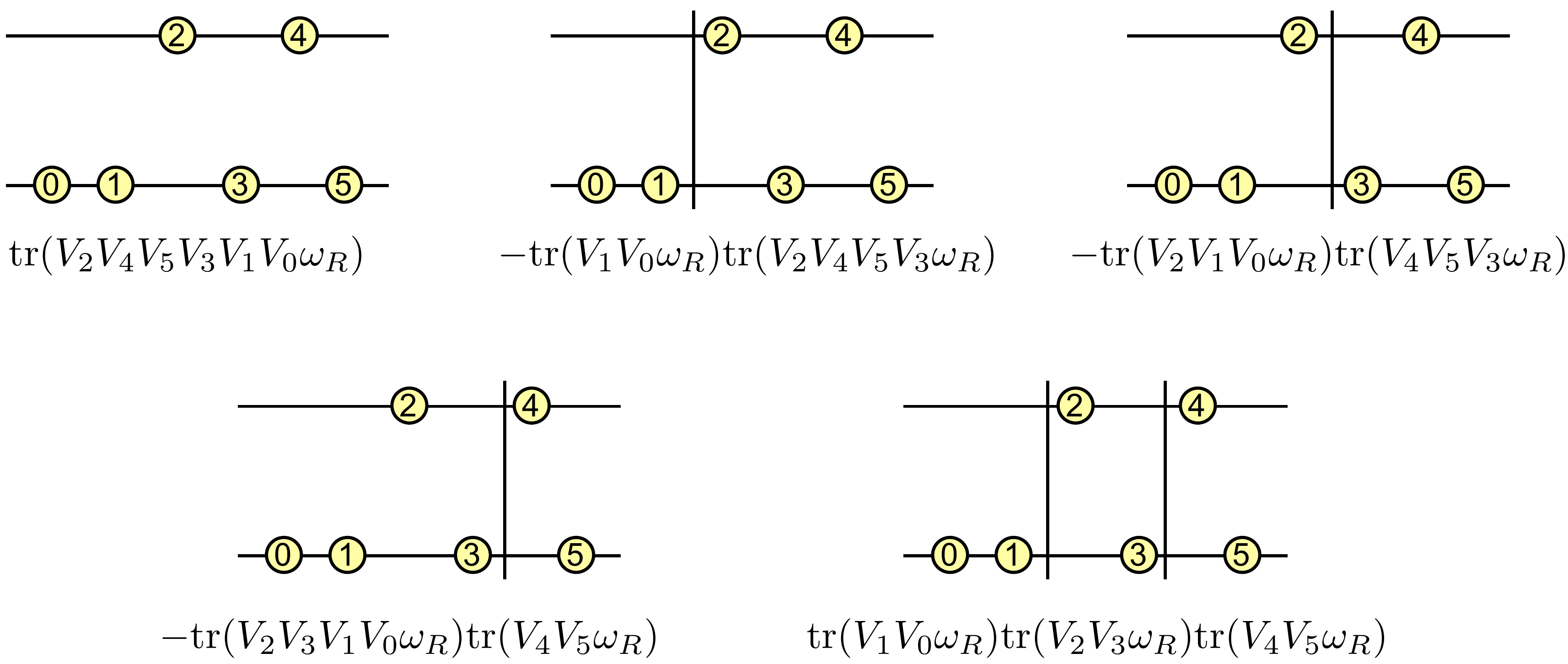}
}
\caption{(Color online) Feynman diagrams of the terms $(-1)^{|d|+1}\prod_{s=1}^{|d|} \tr\!\left( \prod_{k \in d_{s}^A } V(z_k)\omega_R \right)$, for all possible $d \in \mathrm{NC}_5$ and $A=\{2,4\}$.}
\label{fig3} 
\end{figure}
\begin{enumerate}
\item
If $d=(d_1)$, $d_1=(0,1,\dots,n+1)$, we have that $|d|=1$ and the corresponding term  in the sum (\ref{comb1}) is
$$
P\mathcal{L}_{SR}(z_{n+1})\cdots \mathcal{L}_{SR}(z_{0})P.
$$
In this situation all the variables $\{ z_0,z_1, \dots , z_{n+1}\}$ stay together between two projections $P$.
\item
If $d=(d_1,d_2,d_3)$, $d_1=(0,1)$, $d_2=(2,3,4)$, $d_3=(5, \dots,n+1)$, we have that $|d|=3$  and the corresponding term  in the sum (\ref{comb1}) is
\begin{eqnarray*}
\hspace*{-5truemm}&&
(P\mathcal{L}_{SR}(z_{n+1})\cdots \mathcal{L}_{SR}(z_{5})P)\nonumber\\
\hspace*{-5truemm}&&\qquad
{}\times
(P\mathcal{L}_{SR}(z_{4})\mathcal{L}_{SR}(z_{3})\mathcal{L}_{SR}(z_{2})P) 
(P\mathcal{L}_{SR}(z_{1})\mathcal{L}_{SR}(z_{0})P).
\end{eqnarray*}
In this case there are three sets of variables that stay together between two projections $P$: $\{ z_0,z_1 \}$,  $\{ z_2,z_3, z_4 \}$ and $\{z_5, \dots , z_{n+1}\}$.
\end{enumerate}
In general we can say that given $d\in \mathrm{NC}_{n+1} $ the corresponding term in the sum (\ref{comb1}) have $|d|$ sets of variables that stay together between two projections $P$.

In order to obtain a more explicit formula for $K_n(t)$, let us first look at the cases $n=1,2$. 
Put $V_k:=V(z_k)= e^{iz_k H_R}Ve^{-iz_k H_R}$ and $W_k:=W(z_k)=e^{iz_k H_S}We^{-iz_k H_S}$. 
Let $\rho \in \mathcal{T}(\mathcal{H})$ and $P\rho=\sigma \otimes \omega_R$. Then,
\begin{eqnarray*}
\hspace*{-5truemm}&&
K_1(t)\rho  =  \int_{\Delta^{2}(t)}P \mathcal{L}_{SR}(z_{2})Q \mathcal{L}_{SR}(z_1)Q\mathcal{L}_{SR}(z_0)(\sigma \otimes \omega_R) \, \d z_1 \d z_2  \nonumber \\
\hspace*{-5truemm}&&\hphantom{K_1(t)\rho}
= \int_{\Delta^{2}(t)}P \mathcal{L}_{SR}(z_{2})\mathcal{L}_{SR}(z_1)\mathcal{L}_{SR}(z_0)(\sigma \otimes \omega_R) \, \d z_1 \d z_2,
\end{eqnarray*}
where we used the fact that $ \mathrm{NC}_2$ contains a unique element $d=(d_1)$ with $d_1=(0,1,2)$. By a direct computation 
it follows that
\begin{eqnarray}\label{eqn:k1new}
\hspace*{-10truemm}&&
K_1(t)\rho=\int_{\Delta^{2}(t)}\d z_1\d z_2\,[\tr(V_{2}V_{1}V_{0}\omega_R) W_{2}W_{1}W_{0}\sigma \nonumber \\[-3truemm]
\hspace*{-10truemm}&&\hphantom{K_1(t)\rho=\int_{\Delta^{2}(t)}\d z_1\d z_2\,[}
{}-\tr(V_{1}V_{0}\omega_RV_{2})W_{1}W_{0}\sigma W_{2}  \nonumber \\
\hspace*{-10truemm}&&\hphantom{K_1(t)\rho=\int_{\Delta^{2}(t)}\d z_1\d z_2\,[}
{}-\tr(V_{2}V_{0}\omega_RV_{1})W_{2}W_{0} \sigma W_{1}\nonumber\\
\hspace*{-10truemm}&&\hphantom{K_1(t)\rho=\int_{\Delta^{2}(t)}\d z_1\d z_2\,[}
{}+ \tr(V_{0}\omega_R V_{1}V_{2})W_{0} \sigma W_{1}W_{2}\nonumber \\
\hspace*{-8truemm}&&\hphantom{K_1(t)\rho=\int_{\Delta^{2}(t)}\d z_1\d z_2\,[}
{}-\tr( V_{2}V_{1}\omega_R V_{0})W_{2}W_{1}\sigma W_{0}\nonumber\\
\hspace*{-10truemm}&&\hphantom{K_1(t)\rho=\int_{\Delta^{2}(t)}\d z_1\d z_2\,[}
{}+\tr(V_{1}\omega_RV_{0} V_{2})W_{1}\sigma W_{0}W_{2} \nonumber \\
\hspace*{-10truemm}&&\hphantom{K_1(t)\rho=\int_{\Delta^{2}(t)}\d z_1\d z_2\,[}
{}+\tr(V_{2}\omega_RV_{0} V_{1})W_{2}\sigma W_{0} W_{1}\nonumber\\
\hspace*{-10truemm}&&\hphantom{K_1(t)\rho=\int_{\Delta^{2}(t)}\d z_1\d z_2\,[}
{}-\tr(\omega_R V_{0} V_{1}V_{2})\sigma W_{0} W_{1}W_{2}]\otimes \omega_R.
\end{eqnarray} 
Observe that the indices of the elements on the left-hand side of $\sigma$ are always decreasing, while the indices on the right-hand side 
of $\sigma$ are increasing.
Therefore, with each term of the sum (\ref{eqn:k1new}) we can associate two disjoint subsets of $\{0,1,2\}$ (the set of the indices) corresponding to the increasing  and to the decreasing  indices; moreover the sign of each term is determined by the number of  increasing indices.  Therefore,
\begin{eqnarray*}
\hspace*{-7truemm}&&
K_1(t)\rho
=\sum_{A \subset \llbracket 2\rrbracket } (-1)^{|A|} \int_{\Delta^{2}(t)}  \tr\!\left(\mathop{\overleftarrow{\prod}}_{j \in \bar{A}} V_{j}\omega_R \mathop{\overrightarrow{\prod}}_{k \in A}V_k  \right) \mathop{\overleftarrow{\prod}}_{j \in \bar{A}}W_{j}  \sigma \mathop{\overrightarrow{\prod}}_{k \in A}W_k \otimes \omega_R \, \d z \nonumber \\
\hspace*{-7truemm}&&\hphantom{K_1(t)\rho}
= \sum_{A \subset \llbracket 2\rrbracket } (-1)^{|A|} \int_{\Delta^{2}(t)} \tr\!\left(\mathop{\overrightarrow{\prod}}_{k \in A}V_k\mathop{\overleftarrow{\prod}}_{j \in \bar{A}}V_{j}\omega_R   \right) \mathop{\overleftarrow{\prod}}_{j \in \bar{A}}W_{j}\sigma\mathop{\overrightarrow{\prod}}_{k \in A}W_k \otimes \omega_R \, \d z, \end{eqnarray*}
where the cyclic property of the trace was used.

Let us now look at $K_2(t)$. We get $\mathrm{NC}_3 =\{ d, d'\}$, with $d=(d_1)$, $d_1=(0,1,2,3)$, and $d'=(d'_1,d'_2)$, with  $d'_1=(0,1), d'_2=(2,3)$.
Using (\ref{comb1}), one has that
\begin{eqnarray}\label{k2}
\hspace*{-7truemm}&&
K_2(t)\rho
=\int_{\Delta^{3}(t)} \d z_3 \d z_2 \d z_1\,[
P \mathcal{L}_{SR}(z_{3})\mathcal{L}_{SR}(z_2) \mathcal{L}_{SR}(z_1)\mathcal{L}_{SR}(z_0)(\sigma \otimes \omega_R)
  \nonumber \\[-3truemm]
\hspace*{-7truemm}&&\hphantom{K_2(t)\rho
=\int_{\Delta^{3}(t)} \d z_3 \d z_2 \d z_1\,[}
{}-P \mathcal{L}_{SR}(z_{3})\mathcal{L}_{SR}(z_2)P \mathcal{L}_{SR}(z_1)\mathcal{L}_{SR}(z_0)(\sigma \otimes \omega_R)],
\nonumber\\
\hspace*{-7truemm}&&
\end{eqnarray}
and by a direct computation which uses the cyclic property of the trace, one finds that
\begin{eqnarray*}
\hspace*{-7truemm}&&
K_2(t)\rho
\nonumber\\
\hspace*{-7truemm}&&\ \ 
=\int_{\Delta^{3}(t)}\d z\,\Big([ \tr(V_3V_2V_1V_0\omega_R)-\tr(V_3V_2\omega_R)\tr(V_1V_0\omega_R)]W_3W_2 W_1W_0 \sigma   \nonumber \\[-3truemm]
\hspace*{-7truemm}&&\ \ 
\hphantom{=\int_{\Delta^{3}(t)}\d z\,\Big(}
{}-[ \tr(V_3V_2V_1V_0\omega_R)-\tr(V_3V_2\omega_R)\tr(V_1V_0\omega_R)]W_2W_1W_0\sigma W_3 \nonumber \\
\hspace*{-7truemm}&&\ \ 
\hphantom{=\int_{\Delta^{3}(t)}\d z\,\Big(}
{}- [\tr(V_2V_3V_1V_0\omega_R)-\tr(V_2V_3\omega_R)\tr(V_1V_0\omega_R)]W_3W_1W_0\sigma W_2 \nonumber \\
\hspace*{-7truemm}&&\ \ 
\hphantom{=\int_{\Delta^{3}(t)}\d z\,\Big(}
{}+ [\tr(V_2V_3V_1V_0\omega_R)-\tr(V_2V_3\omega_R)\tr(V_1V_0\omega_R)]W_1W_0\sigma W_2W_3 \nonumber \\
\hspace*{-7truemm}&&\ \ 
\hphantom{=\int_{\Delta^{3}(t)}\d z\,\Big(}
{}- [\tr(V_1V_3V_2V_0\omega_R)-\tr(V_3V_2\omega_R)\tr(V_1V_0\omega_R)]W_3W_2W_0\sigma W_1 \nonumber \\
\hspace*{-7truemm}&&\ \ 
\hphantom{=\int_{\Delta^{3}(t)}\d z\,\Big(}
{}+ [\tr(V_1V_3V_2V_0\omega_R)-\tr(V_3V_2\omega_R)\tr(V_1V_0\omega_R)]W_2W_0\sigma W_1W_3 \nonumber \\
\hspace*{-7truemm}&&\ \ 
\hphantom{=\int_{\Delta^{3}(t)}\d z\,\Big(}
{}+ [\tr(V_1V_2V_3V_0\omega_R)-\tr(V_2V_3\omega_R)\tr(V_1V_0\omega_R)]W_3W_0\sigma W_1W_2 \nonumber \\
\hspace*{-7truemm}&&\ \ 
\hphantom{=\int_{\Delta^{3}(t)}\d z\,\Big(}
{}- [\tr(V_1V_2V_3V_0\omega_R)-\tr(V_2V_3\omega_R)\tr(V_1V_0\omega_R)]W_0\sigma W_1W_2W_3 \nonumber \\
\hspace*{-7truemm}&&\ \ 
\hphantom{=\int_{\Delta^{3}(t)}\d z\,\Big(}
{}- [\tr(V_0V_3V_2V_1\omega_R)-\tr(V_3V_2\omega_R)\tr(V_0V_1\omega_R)]W_3W_2W_1\sigma W_0 \nonumber \\
\hspace*{-7truemm}&&\ \ 
\hphantom{=\int_{\Delta^{3}(t)}\d z\,\Big(}
{}+ [\tr(V_0V_3V_2V_1\omega_R)-\tr(V_3V_2\omega_R)\tr(V_0V_1\omega_R)]W_2W_1\sigma W_0W_3 \nonumber \\
\hspace*{-7truemm}&&\ \ 
\hphantom{=\int_{\Delta^{3}(t)}\d z\,\Big(}
{}+ [\tr(V_0V_2V_3V_1\omega_R)-\tr(V_2V_3\omega_R)\tr(V_0V_1\omega_R)]W_3W_1\sigma W_0W_2 \nonumber \\
\hspace*{-7truemm}&&\ \ 
\hphantom{=\int_{\Delta^{3}(t)}\d z\,\Big(}
{}- [\tr(V_0V_2V_3V_1\omega_R)-\tr(V_2V_3\omega_R)\tr(V_0V_1\omega_R)]W_1\sigma W_0W_2W_3 \nonumber \\
\hspace*{-7truemm}&&\ \ 
\hphantom{=\int_{\Delta^{3}(t)}\d z\,\Big(}
{}+ [\tr(V_0V_1V_3V_2\omega_R)-\tr(V_3V_2\omega_R)\tr(V_0V_1\omega_R)]W_3W_2\sigma W_0W_1 \nonumber \\
\hspace*{-7truemm}&&\ \ 
\hphantom{=\int_{\Delta^{3}(t)}\d z\,\Big(}
{}- [\tr(V_0V_1V_3V_2\omega_R)-\tr(V_3V_2\omega_R)\tr(V_0V_1\omega_R)]W_2\sigma W_0W_1W_3 \nonumber \\
\hspace*{-7truemm}&&\ \ 
\hphantom{=\int_{\Delta^{3}(t)}\d z\,\Big(}
{}- [\tr(V_0V_1V_2V_3\omega_R)-\tr(V_2V_3\omega_R)\tr(V_0V_1\omega_R)]W_3\sigma W_0W_1W_2 \nonumber \\[-1truemm]
\hspace*{-7truemm}&&\ \ 
\hphantom{=\int_{\Delta^{3}(t)}\d z\,\Big(}
{}+[\tr(V_0V_1V_2V_3\omega_R)-\tr(V_2V_3\omega_R)\tr(V_0V_1\omega_R)] \sigma W_0W_1W_2W_3 \Big)
\nonumber \\
\hspace*{-7truemm}&&\hspace*{117truemm}
{}\otimes \omega_R.
\end{eqnarray*}
The only difference with the case $n=1$ consists in the content of the square brackets.
There are two terms in the first one all the variables stay together, similarly to the case $n=1$, while in the second one there are two sets of variables that stay together, $\{z_2,z_3\}$ and $\{z_0,z_1\}$.  
In each square bracket the first  term comes from the first line of~(\ref{k2}), namely from the partition $d$, while the second term comes from the second line, namely from the partition~$d'$.

Generalizing these considerations to an arbitrary $n$ it can be proved by induction that  $K_n(t)\rho$ can be written as follows:
\begin{eqnarray*}
\hspace*{-6truemm}&&
K_n(t)\rho=\sum_{A \subset \llbracket n+1\rrbracket }  (-1)^{|A|} \int_{\Delta^{n+1}(t)}\d z\mathop{\overleftarrow{\prod}}_{j \in \bar{A}}W(z_{j})\sigma\mathop{\overrightarrow{\prod}}_{k \in A}W(z_k) \otimes \omega_R   \nonumber \\
\hspace*{-6truemm}&&\hspace*{54truemm}
{}\times\sum_{d \in \mathrm{NC}_{n+1}} (-1)^{|d|+1}\prod_{s=1}^{|d|} \tr\!\left(\prod_{k \in d_{s}^A } V(z_k)\omega_R \right)\nonumber \\
\hspace*{-6truemm}&&\hphantom{K_n(t)\rho}
=\sum_{A \subset \llbracket n+1\rrbracket }  (-1)^{|A|} \int_{\Delta^{n+1}(t)}\d z \,  \mathcal{G}_n(A,z)\mathop{\overleftarrow{\prod}}_{j \in \bar{A}}W(z_{j})  \sigma\mathop{\overrightarrow{\prod}}_{k \in A} W(z_k) \otimes \omega_R,  
\end{eqnarray*}
where $d_s^A$ is the rearrangement of $d_s$ by $A$,  as defined in Definition~\ref{defn:dA}, and 
$$
\mathcal{G}_n(A,z)= \sum_{d \in \mathrm{NC}_{n+1}} (-1)^{|d|+1} \prod_{s=1}^{|d|} \tr\!\left(\prod_{k \in d_s^A } V(z_k)\omega_R \right).
$$
\end{pf}

\section{Main result}\label{sec:mainth}
By using Davies' abstract result and the above diagrammatic expansion, we will prove the existence of the limit dynamics (\ref{eqn:limit dynamics}) for a finite-dimensional system $S$ weakly coupled to a generic reservoir $R$, when the coupling operator $V$ and the reference state $\omega_R$ satisfy Assumptions A and some additional suitable assumptions. 

First of all, let us recall when a state is \emph{mixing}.
Let $\omega \in \mathcal{T}(\mathcal{H}_R)$ be positive and normalized. We say that $\omega$ is mixing if for any bounded operators $A$ and $B$ on $\mathcal{H}_R$ one has
\begin{equation}
\lim_{t \to +\infty} \tr(A(t)B \omega)=\tr(A\omega)\tr(B\omega),
\label{eq:mixing}
\end{equation}
where $A(t)=e^{it\underline{\mathcal{L}}_R} A = e^{itH_R}Ae^{-itH_R}$. This can be proved to be equivalent to the condition~\cite{ref:JaksicPillet}
\begin{equation}
\mathop{\textrm{w-lim}}_{t \to +\infty} e^{-itH_R}=P_{\Omega_R},
\label{eq:weakdecay}
\end{equation}
where $\mathop{\textrm{w-lim}}$ denotes the weak limit, and $\omega_R=P_{\Omega_R}$ is the rank-one projection associated with the reference state of the reservoir  $\Omega_R$.

In order to prove the convergence of van Hove's limit we will need an interaction $V$ whose correlations are decaying sufficiently fast. Remember the assumption~(\ref{eq:meanV0}), $\tr(V \omega_R)= \langle \Omega_R | V\Omega_R\rangle=0$, which means that the vector $v=V\Omega_R$ is orthogonal to the reference state $\Omega_R$. Thus, by~(\ref{eq:weakdecay}) we have that the two-point correlation function decays,
\begin{equation}
\tr(V(t) V \omega_R) = \langle v| e^{-it H_R} v\rangle \to 0,
\end{equation}
as $t\to+\infty$. We will require that it decays fast enough, such that it is integrable.

In fact,  we will need a stronger mixing property,  given by the following conditions on the $n$-point correlation functions.
\begin{defn}\label{defn:gapcond}
The triple $(H_R, V, \omega_R)$ has a \emph{clustering property} if there exists a function $f : \RM \to \RM$ that satisfies the following conditions:
\begin{itemize}
\item $f$ is non-negative and
$$
\|f\|_{1,\epsilon}:=\int_{\RM}f(s)(1+|s|)^\epsilon \, \d s < + \infty,  
$$
for some $0<\epsilon<1$.
\item There exists $C>0$ such that for all $n\geq 1$ and for all $A \subset \llbracket n+1\rrbracket $ it results that
$$
| \mathcal{G}_{n}(A,z)| \leq \frac{C^{n+2}}{\left[ \frac{n}{2}\right]!} \sum_{p \in \mathcal{S}_{n}'}  \prod_{l=0}^{\left[ \frac{n+1}{2}\right]}   f(z_{p(l)}-z_{p(l+1)}),
$$
where $\mathcal{S}_{n}'$ denotes the set of all the permutations $p$ of $\{0,1, \dots, n+1\}$ such that 
$|p(1)-p(0)|\geq2$ (gapped permutations).
\end{itemize}
\end{defn}

Roughly speaking, the clustering property bounds the 
$(n+2)$-point connected correlation function $\mathcal{G}_{n}(A,z)$ by the product of $[\frac{n+1}{2}]+1$ two-point functions, where at least one of the pairs of times is taken at two nonconsecutive times. It is related to the strong mixing property
\begin{equation}
\lim_{t \to +\infty} \tr(A B(t) C \omega)=\tr(AC\omega)\tr(B\omega),
\label{eqn:StrongMixing}
\end{equation}
which obviously implies~(\ref{eq:mixing}). 
Indeed, under the strong mixing condition (\ref{eqn:StrongMixing}), one can show that the $(n+2)$-point correlation $\mathcal{G}_{n}(A,z)$ decays as the separation of any pair of consecutive times increases,
\begin{equation}
\mathcal{G}_{n}(A,z)\to0
\text{as}
z_k-z_{k-1}\to+\infty
\end{equation}
for $k=1,\ldots,n+1$, which implies that it decays as the separation of any pair of times increases,
\begin{equation}
\mathcal{G}_{n}(A,z)\to0
\text{as}
z_k-z_j\to+\infty,
\end{equation}
for $k,j=0,1,\dots,n+1$, $k>j$.
Note also that the strong mixing property and the clustering property cannot hold if $V$ is a finite-rank operator. One can argue that this is physically sensible, because in such a case the system would  see, through $V$,  an effective finite-dimensional reservoir.

\paragraph*{\emph{ASSUMPTIONS B:}}
\begin{enumerate}
\item The correlation function $\varphi: \RM \to \CM$, such that for all $t \in \RM$
$$
\varphi(t):=
\tr(V(t)V\omega_R),
$$
is in $L^1(\RM)$, namely
$$
\| \varphi \|_{1}:=\int_{\RM} |\varphi(t)| \, \d t < +\infty.
$$ 
\item The triple $(H_R,V,\omega_R)$ has a clustering property.
\end{enumerate}

Notice that  Assumption~B1 implies a mixing property only on the two-point correlation function of the observable $V$. In general there can exist a pair of observables $A$ and $B$, different from $V$, which do not satisfy~(\ref{eq:mixing}), whence mixing is neither sufficient nor necessary for this Assumption~B1 to hold.

\begin{thm}\label{thm:mainth}
Let Assumptions A and B hold, and let $K^\lambda(\tau)$ be defined by~(\ref{eqn:Klambdatau}). Then
one gets
\begin{equation}\label{eqn:KlambdatoK}
\lim_{\lambda \to 0} K^\lambda(\tau)=\bm{K}
\end{equation}
uniformly in $\tau \in [\tau_0, \tau_1]$, $0<\tau_0<\tau_1$,
where the bounded operator $\bm{K}$ acting on $\mathcal{T}(\mathcal{H})$ is given by~(\ref{eqn:def_K}).  
Moreover,
\begin{equation}\label{eqn:limiting dynamicsY}
\lim_{\lambda \to 0} U^{\lambda}(\tau)=e^{-\tau \bm{K}^{\natural}} P
\end{equation}
uniformly in $\tau \in [0, \tau_1], \tau_1 >0 $, where 
\begin{equation}
\bm{K}^{\natural}=\lim_{T \to + \infty}\frac{1}{T} \int_0^T e^{is\mathcal{L}_S}\bm{K}e^{-is\mathcal{L}_S}\,\d s
\end{equation}
is Davies' spectral average of $\bm{K}$.
\end{thm}

We split the proof of Theorem~\ref{thm:mainth} into two propositions.
\begin{prop}\label{lemmaKn}
If Assumptions A and B hold, then for all  $t >0$ one has that  for all $n \geq 1$ the operator $K_n(t)$ given by~(\ref{defn:Kntabstract}) satisfies the bound
\begin{equation}\label{eqn:estKn}
\|K_n(t)\|\leq  c_n t^{[n/2]},
\end{equation}
where 
\begin{equation}
c_n= \frac{(2 C\|W\|)^{n+2}}{\left[ \frac{n}{2}\right]!}\|f\|_1^{\left[ \frac{n+1}{2}\right]+1}  ,
\end{equation}
with $C$ and $f$ as in Definition~\ref{defn:gapcond}.
\end{prop}
\begin{pf} 
By Theorem~\ref{thm:Kncomb} we have that, for all $n \geq 1$ and $\rho \in \mathcal{T}(\mathcal{H})$, $\|\rho\|=1$,
$$
 \|K_n(t) \rho \|\leq \|W\|^{n+2}     \sum_{A \subset  \llbracket n+1\rrbracket  } \int_{\Delta^{n+1}(t)} |\mathcal{G}_n(A,z)|  \, \d z.
$$
Moreover, since $(H_R,V, \omega_R)$ has a clustering property, according to Definition~\ref{defn:gapcond} we have that, for all $A \subset \llbracket n+1\rrbracket $,
\begin{eqnarray*}
\hspace*{-5truemm}&&
\int_{\Delta^{n+1}(t)}|\mathcal{G}_n(A,z)|\,\d z
\leq   \frac{C^{n+2}}{\left[ \frac{n}{2}\right]!} \sum_{p \in \mathcal{S}_{n}'} \int_{\Delta^{n+1}(t)} \prod_{l=0}^{\left[ \frac{n+1}{2}\right]}   f(z_{p(l)}-z_{p(l+1)}) \nonumber \\
\hspace*{-5truemm}&&\hphantom{\int_{\Delta^{n+1}(t)}|\mathcal{G}_n(A,z)|\,\d z}
\leq\frac{C^{n+2}}{\left[ \frac{n}{2}\right]!}  \int_{[0,t]^{n+1}} \prod_{l=0}^{\left[ \frac{n+1}{2}\right]}   f(z_{l}-z_{l+1}) \, \d z.
\end{eqnarray*}
Therefore
\begin{eqnarray*}
\hspace*{-5truemm}&&
\|K_n(t)\rho\|
\leq\frac{(2 C\|W\|)^{n+2}}{\left[ \frac{n}{2}\right]!}  \int_{[0,t]^{n+1}} \prod_{l=0}^{\left[ \frac{n+1}{2}\right]}   f(z_{l}-z_{l+1}) \, \d z   \nonumber \\
\hspace*{-5truemm}&&\hphantom{\|K_n(t)\rho\|}
\leq\frac{(2 C\|W\|)^{n+2}}{\left[ \frac{n}{2}\right]!} \|f\|_1^{\left[ \frac{n+1}{2}\right]+1}    t^{[n/2]},
\end{eqnarray*}
and this proves (\ref{eqn:estKn}). 
\end{pf}

\begin{prop}\label{lemmaKneps}
Let Assumptions A and B hold, and let $K_n(t)$ acting on $\mathcal{T}(\mathcal{H})$ be given by~(\ref{defn:Kntabstract}). Then, we have that, for all  $t >0$ and $m \geq 1$,
\begin{equation}
\label{eqn:estKn_eps}
 \|K_{2m}(t)\|  \leq   d_m   t^{m-\epsilon},                                         
\end{equation}
where 
\begin{equation}
d_m=\frac{(2 C\|W\|)^{2m+2}}{m!} (2m+2)! \| f\|_{1,\epsilon}  \xi^{(\epsilon)}_m,
\end{equation}
\begin{equation}
\xi_m^{(\epsilon)}
=\max_{\stackrel{k>i+1}{k,i=0,\ldots, m+1}}
\frac{
(k-i-1-\epsilon)^{k-i-1-\epsilon}
(m-k+i+1)^{m-k+i+1}
}{(m-\epsilon)^{m-\epsilon}(k-i-1)!(m-k+i+1)!},
\label{eqn:XiProp2}
\end{equation}
with the norm  $\|f\|_{1,\epsilon}$ given in Definition~\ref{defn:gapcond}.
\end{prop}
\begin{pf}
By Theorem~\ref{thm:Kncomb} we have that, for all $m \geq 1$ and $\rho \in \mathcal{T}(\mathcal{H})$, $\|\rho\|=1$,
$$
 \|K_{2m}(t) \rho \|\leq \|W\|^{2m+2}     \sum_{A \subset  \llbracket 2m+1\rrbracket } \int_{\Delta^{2m+1}(t)} |\mathcal{G}_{2m}(A,z)|  \, \d z.
$$
Moreover, since $(H_R,V, \omega_R)$ is a clustering triple, according to Definition~\ref{defn:gapcond} we have that, for all $A \subset \llbracket 2m+1\rrbracket$,
$$
\int_{\Delta^{2m+1}(t)} |\mathcal{G}_{2m}(A,z)|\,  \d z \leq   \frac{C^{2m+2}}{m!} \sum_{p \in \mathcal{S}_{n}'} \int_{\Delta^{2m+1}(t)} \prod_{l=0}^{m}   f(z_{p(l)}-z_{p(l+1)})\,  \d z. 
$$
Notice that, for all $p \in \mathcal{S}_{n}'$,
$$
 \int_{\Delta^{2m+1}(t)} \prod_{l=0}^{m}   f(z_{p(l)}-z_{p(l+1)}) \, \d z \leq \|f\|_1^{m}  \int_{\Delta^{m+1}(t)}    f(z_{k}-z_{i}) \, \d z
$$
for some $k,i \in \{0,1, \dots, m+1\}$ with $|k-i|>1$. We distinguish two cases: if $k>i+1$, by Lemma~\ref{lemma:simplessoind} given in the Appendix we have that
\begin{equation}\label{eqn:estk2mp3}
\int_{\Delta^{m+1}(t)} f(z_k-z_i) \, \d z \leq  \| f\|_{1,\epsilon} \xi^{(\epsilon)}_m t^{m-\epsilon},
\end{equation}
with $\xi_m^{(\epsilon)}$ given in (\ref{eqn:XiProp2}).
If $i>k+1$, then 
$$
\int_{\Delta^{m+1}(t)} f(z_k-z_i) \, \d z =\int_{\Delta^{m+1}(t)} \tilde{f}(z_i-z_k) \, \d z, 
$$
where $\tilde{f}(x)=f(-x)$.  Since $\|\tilde{f}\|_{1,\epsilon}=\|f\|_{1,\epsilon}$, by 
Lemma~\ref{lemma:simplessoind} given in the Appendix we again have (\ref{eqn:estk2mp3}).
Therefore,
$$
 \|K_{2m}(t) \rho \|\leq \frac{(2 C\|W\|)^{2m+2}}{m!} (2m+2)! \| f\|_1^m\| f\|_{1,\epsilon} \xi^{(\epsilon)}_m t^{m-\epsilon},
$$
which proves (\ref{eqn:estKn_eps}). 
Note that this estimate fails if we drop the condition of ``gapped permutations'' in the definition of the clustering property in Definition~\ref{defn:gapcond}, since Lemma~\ref{lemma:simplessoind} requires a gap.
\end{pf}

Armed with Propositions~\ref{lemmaKn} and~\ref{lemmaKneps}, we can at last conclude the proof of Theorem~\ref{thm:mainth}.

\bigskip
\begin{pf}(Theorem~\ref{thm:mainth})
First we prove (\ref{eqn:KlambdatoK}). We observe that for all $\sigma \otimes \omega_R \in P\mathcal{T}(\mathcal{H})$,
\begin{eqnarray*}
\hspace*{-8truemm}&&
\bm{K}(\sigma \otimes \omega_R )\nonumber \\
\hspace*{-8truemm}&&\quad
=\int_{0}^{+\infty}\d z\,\Bigl(
\tr(V(z)V\omega_R)[W(z),W \sigma]-\tr(VV(z)\omega_R)[W(z),\sigma W]\Bigr)\otimes \omega_R\nonumber \\
\hspace*{-8truemm}&&\quad
=\int_{0}^{+\infty}\d z\,\Bigl(
\varphi(z)[W(z),W \sigma]- \varphi(-z)[W(z),\sigma W]\Bigr) \otimes \omega_R. \nonumber
\end{eqnarray*}
Therefore it results that
$$
\|\bm{K}\|\leq 4 \|W\|^2 
\| \varphi \|_1 < + \infty.
$$
Combining Proposition~\ref{lemmaKn}, Proposition~\ref{lemmaKneps}  and Theorem~\ref{theorem:davies},  we obtain (\ref{eqn:limiting dynamicsY}). 
\end{pf}

\section*{Appendix A}
\def\theequation{A.\arabic{equation}}
\setcounter{equation}{0}
\label{appendix}
Here we prove a technical lemma needed in the proof of the main Theorem~\ref{thm:mainth}.
\begin{lem}\label{lemma:simplessoind}
Let $g \in L^1(\RM)$, then for all $m \geq 1$, for all $t >0$ and for all $k,i \in \{0, \dots, m+1\}$, $k >i$, it results that
\begin{equation}\label{en:simplex equality}
\int_{\Delta^{m+1}(t)} g(z_k-z_i) \,\d z=\int_{0}^t  g(s) \frac{s^{k-i-1}}{(k-i-1)!}\frac{(t-s)^{m-k+i+1}}{(m-k+i+1)!}\, \d s,
\end{equation}
where $z_0:=0$.
Moreover, if $g \geq 0$ and
$$
\|g\|_{1,\epsilon}:=\int_{\RM}g(z) (1+|z|)^\epsilon \,\d z < +\infty, \text{for some $\epsilon >0$,}
$$
then for $k>i+1$,
\begin{equation}\label{eqn:eps_estimate}
\int_{\Delta^{m+1}(t)} g(z_k-z_i) \, \d z \leq \|g \|_{1,\epsilon} \xi_m^{(\epsilon)} t^{m-\epsilon},
\end{equation}
where
\begin{equation}
\xi_m^{(\epsilon)}
=\max_{\stackrel{k>i+1}{k,i=0,\ldots, m+1}}
\frac{
(k-i-1-\epsilon)^{k-i-1-\epsilon}
(m-k+i+1)^{m-k+i+1}
}{(m-\epsilon)^{m-\epsilon}(k-i-1)!(m-k+i+1)!}.
\label{eq:xieps}
\end{equation}
\end{lem}
\begin{pf}
We start with the proof of (\ref{en:simplex equality}). 
Let us first look at a simple case, with $k=2$ and $i=1$ for $m=1$,
\begin{eqnarray*}\label{eqn:M1}
\hspace*{-5truemm}&&
\int_{\Delta^{2}(t)} g(z_2-z_1)\,  \d z
= \int_{0}^t \d z_2 \int_{0}^{z_2} \d z_1 \,  g(z_2-z_1)
\nonumber\\
\hspace*{-5truemm}&&\hphantom{\int_{\Delta^{2}(t)} g(z_2-z_1)\,\d z}
=  \int_{0}^t \d z_1 \int_{z_1}^{t} \d z_2 \,  g(z_2-z_1)  \nonumber \\
\hspace*{-5truemm}&&\hphantom{\int_{\Delta^{2}(t)} g(z_2-z_1)\,\d z}
=  \int_{0}^t \d z_1 \int_0^{t-z_1} \d z_2 \,  g(z_2)  \nonumber \\
\hspace*{-5truemm}&&\hphantom{\int_{\Delta^{2}(t)} g(z_2-z_1)\,\d z}
=   \int_{0}^t\d s\,g(s) (t-s).    
\end{eqnarray*}
By generalizing this strategy, 
we manipulate the integral for $m+1=k>i\ge1$ as
\begin{eqnarray*}
\hspace*{-5truemm}&&
\int_{\Delta^k(t)}g(z_k-z_i)\,\d z
\nonumber\\
\hspace*{-5truemm}&&\quad
=\int_{0}^t\d z_k\int_{0}^{z_k}\d z_{k-1}\cdots\int_0^{z_2}\d z_1\,g(z_k-z_i)
\nonumber\\
\hspace*{-5truemm}&&\quad
=\int_{0}^t\d z_k\int_{0}^{z_k}\d z_{k-1}\cdots\int_0^{z_{i+1}}\d z_i\,g(z_k-z_i)\frac{z_i^{i-1}}{(i-1)!}
\nonumber\\
\hspace*{-5truemm}&&\quad
=\int_{0}^t\d z_i\int_{z_i}^t\d z_{i+1}\cdots\int_{z_{k-1}}^t\d z_k\,g(z_k-z_i)\frac{z_i^{i-1}}{(i-1)!}
\nonumber\\
\hspace*{-5truemm}&&\quad
=\int_{0}^t\d z_i\int_0^{t-z_i}\d z_{i+1}\int_{z_{i+1}}^{t-z_i}\d z_{i+2}\cdots\int_{z_{k-1}}^{t-z_i}\d z_k\,g(z_k)\frac{z_i^{i-1}}{(i-1)!}
\nonumber\\
\hspace*{-5truemm}&&\quad
=\int_{0}^t\d z_i\int_0^{z_i}\d z_{i+1}\int_{z_{i+1}}^{z_i}\d z_{i+2}\cdots
\int_{z_{k-1}}^{z_i}\d z_k\,g(z_k)\frac{(t-z_i)^{i-1}}{(i-1)!}
\nonumber\\
\hspace*{-5truemm}&&\quad
=\int_{0}^t\d z_i\,\frac{(t-z_i)^{i-1}}{(i-1)!}
\int_0^{z_i}\d z_k\,g(z_k)\int_0^{z_k}\d z_{k-1}
\cdots\int_0^{z_{i+2}}\d z_{i+1}
\nonumber\\
\hspace*{-5truemm}&&\quad
=\int_{0}^t\d z_i\,\frac{(t-z_i)^{i-1}}{(i-1)!}
\int_0^{z_i}\d s\,g(s)\frac{s^{k-i-1}}{(k-i-1)!}
\nonumber\\
\hspace*{-5truemm}&&\quad
=\int_0^t\d s\,g(s)\frac{s^{k-i-1}}{(k-i-1)!}\int_s^t\d z_i\,\frac{(t-z_i)^{i-1}}{(i-1)!}
\nonumber\\
\hspace*{-5truemm}&&\quad
=\int_0^t\d s\,g(s)\frac{s^{k-i-1}}{(k-i-1)!}\frac{(t-s)^i}{i!}.
\end{eqnarray*}
Note that this final formula works also for $k>i=0$. 
Then, for $m+1\ge k>i\ge0$, we have
\begin{eqnarray*}
\hspace*{-5truemm}&&
\int_{\Delta^{m+1}(t)} g(z_k-z_i)\,  \d z
\nonumber\\
\hspace*{-5truemm}&&\quad
= \int_{0}^t \d z_{m+1} \int_{0}^{z_{m+1}} \d z_m\cdots\int_0^{z_2}\d z_1 \,  g(z_k-z_i)
\nonumber\\
\hspace*{-5truemm}&&\quad
= \int_{0}^t \d z_{m+1} \int_{0}^{z_{m+1}} \d z_m\cdots
\int_0^{z_{k+1}}\d s\,g(s)\frac{s^{k-i-1}}{(k-i-1)!}\frac{(z_{k+1}-s)^i}{i!}
\nonumber\\
\hspace*{-5truemm}&&\quad
= \int_0^t\d s\,g(s)\frac{s^{k-i-1}}{(k-i-1)!}
\int_s^t \d z_{k+1}\,\frac{(z_{k+1}-s)^i}{i!}
\int_{z_{k+1}}^t\d z_{k+2}\cdots\int_{z_m}^t\d z_{m+1}
\nonumber\\
\hspace*{-5truemm}&&\quad
= \int_0^t\d s\,g(s)\frac{s^{k-i-1}}{(k-i-1)!}
\int_s^t \d z_{k+1}\,\frac{(z_{k+1}-s)^i}{i!}
\frac{(t-z_{k+1})^{m-k}}{(m-k)!}
\nonumber\\
\hspace*{-5truemm}&&\quad
= \int_0^t\d s\,g(s)
\frac{s^{k-i-1}}{(k-i-1)!}
\frac{(t-s)^{m-k+i+1}}{(m-k+i+1)!},
\end{eqnarray*}
which is (\ref{en:simplex equality}).

Now we prove (\ref{eqn:eps_estimate}) using (\ref{en:simplex equality}).
\begin{eqnarray*}
\hspace*{-5truemm}&&
\int_{\Delta^{m+1}(t)} g(z_k-z_i) \, \d z \nonumber \\
\hspace*{-5truemm}&&\qquad
= \int_{0}^t  g(s) \frac{s^{k-i-1}}{(k-i-1)!} \frac{(t-s)^{m-k+i+1}}{(m-k+i+1)!}\, \d s\nonumber \\
\hspace*{-5truemm}&&\qquad
= \int_{0}^t g(s)(1+s)^\epsilon  \frac{1}{(1+s)^\epsilon} \frac{s^{k-i-1}}{(k-i-1)!}\frac{(t-s)^{m-k+i+1}}{(m-k+i+1)!} \,  \d s \nonumber \\
\hspace*{-5truemm}&&\qquad
\leq   \|g\|_{1,\epsilon} \max_{s\in [0,t]}\frac{1}{(1+s)^\epsilon} \frac{s^{k-i-1}}{(k-i-1)!}\frac{(t-s)^{m-k+i+1}}{(m-k+i+1)!}\nonumber \\
\hspace*{-5truemm}&&\qquad
\leq   \|g\|_{1,\epsilon} \max_{s\in [0,t]}\frac{s^{k-i-1-\epsilon}}{(k-i-1)!}\frac{(t-s)^{m-k+i+1}}{(m-k+i+1)!}\nonumber \\
\hspace*{-5truemm}&&\qquad
=   \|g\|_{1,\epsilon}
\frac{
(k-i-1-\epsilon)^{k-i-1-\epsilon}
(m-k+i+1)^{m-k+i+1}
}{(m-\epsilon)^{m-\epsilon}(k-i-1)!(m-k+i+1)!}
t^{m-\epsilon}
\nonumber \\
\hspace*{-5truemm}&&\qquad
\leq  \|g\|_{1,\epsilon}\xi_m^{(\epsilon)}t^{m-\epsilon},
\end{eqnarray*}
which is~(\ref{eqn:eps_estimate}), where $\xi_m^{(\epsilon)}$ is given by~(\ref{eq:xieps})
\end{pf}

\section*{Acknowledgments}
This work was partially supported by INFN through the project ``QUANTUM", by the Italian National Group of Mathematical Physics (GNFM-INdAM), and by the Top Global University Project from the Ministry of Education, Culture, Sports, Science and Technology (MEXT), Japan.
KY was supported by the Grant-in-Aid for ScientificResearch (C) (No.~26400406) from the Japan Society for the Promotion of Science (JSPS) and by the Waseda University Grant for Special Research Projects (No.~2017K-236).
ML  was  supported by Cohesion and Development Fund 2007-2013 - APQ Research Puglia Region ``Regional program supporting smart specialization and social and environmental sustainability - FutureInResearch''.

\end{document}